\documentclass[twocolumn,reprint,aps,prd,superscriptaddress,nofootinbib]{revtex4-1}
\usepackage[utf8]{inputenc}
\usepackage[english]{babel}
\usepackage[toc,page]{appendix}
\usepackage{amsmath,bm}
\usepackage{svg} 
\usepackage{mathrsfs}
\usepackage{amsfonts}
\usepackage{amssymb}
\usepackage{bbm}
\usepackage{graphicx}
\usepackage{xcolor}
\usepackage{hyperref}
\usepackage{physics}
\usepackage{braket}
\usepackage{bm}
\usepackage{mathtools}
\usepackage{float}
\usepackage{tikz} 
\hypersetup{
    pdftitle={},
    colorlinks=true,
    linkcolor=blue,
    citecolor=blue,
    filecolor=blue,
    urlcolor=blue
}
\usepackage{cleveref}

\begin{document}
\title{Entropy transport in closed quantum many-body systems far from equilibrium}
\author{Jendrik Marijan}
\email{marijan@thphys.uni-heidelberg.de}
\affiliation{Institut für Theoretische Physik, Universität Heidelberg, Philosophenweg 12/16, 69120 Heidelberg, Germany}
\author{Helmut Strobel}
\affiliation{Kirchhoff Institut für Physik, Universität Heidelberg, Im Neuenheimer Feld 227, 69120 Heidelberg, Germany}
\author{Markus K.\ Oberthaler}
\affiliation{Kirchhoff Institut für Physik, Universität Heidelberg, Im Neuenheimer Feld 227, 69120 Heidelberg, Germany}
\author{Jürgen Berges}
\affiliation{Institut für Theoretische Physik, Universität Heidelberg, Philosophenweg 12/16, 69120 Heidelberg, Germany}
\begin{abstract}
We investigate entropy transport for universal scaling phenomena in closed quantum many-body systems far from equilibrium. From spatially resolved experimental data of a spinor Bose gas, we demonstrate that entropy decreases on long-distance scales while it increases at short distances. A dynamical separation of scales leads to macrophysics with long-range order, which is insensitive to the highly entropic microphysical processes. 
Since the total von Neumann entropy is conserved on a fundamental level for the quantum system, our analysis reveals a reciprocal connection between the emergence of macroscopic structure and microscopic disorder. To illustrate the scope of this connection, we exemplify the universal phenomenon also in a relativistic quantum field theory calculation from first principles, which is relevant for particle physics and early-universe cosmology.  
\end{abstract}
\maketitle
\section{Introduction and overview}\label{sec:introduction}
\begin{figure*}
    \includegraphics[width=0.98\linewidth]{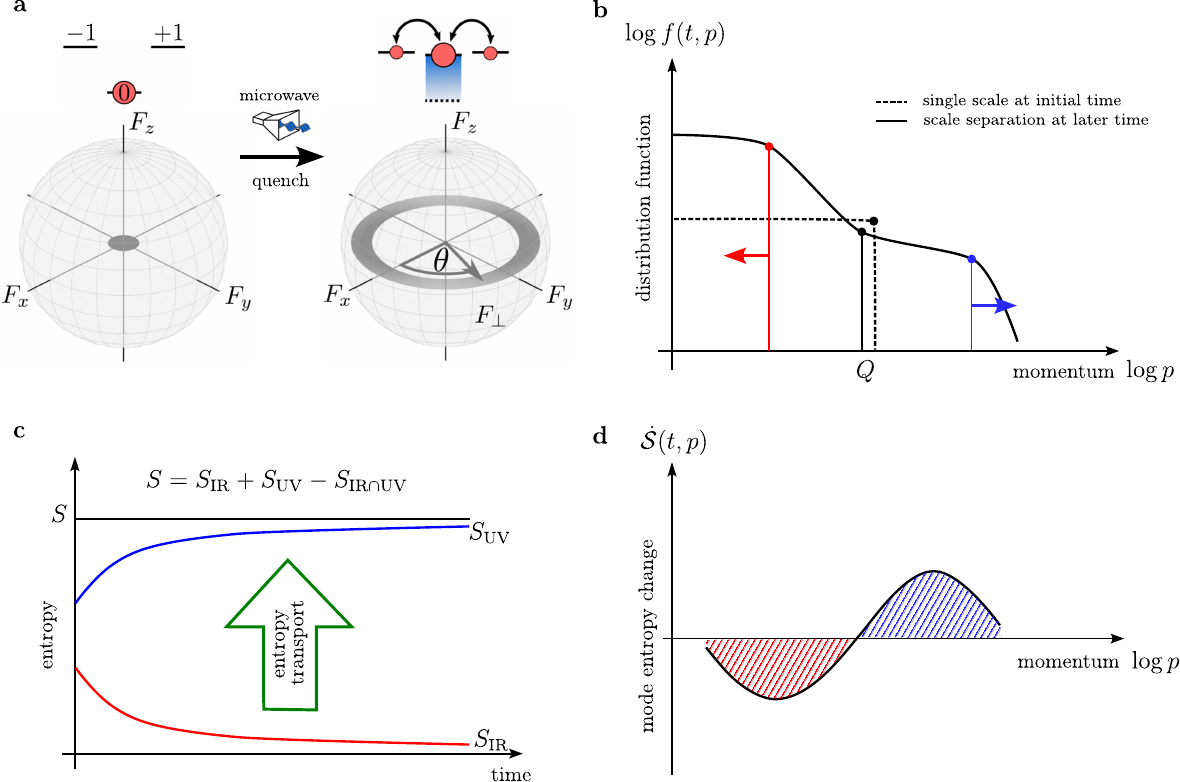}
    \caption{Illustration: \textbf{a) Experimental scheme} for the spinor Bose gas with three hyperfine states. After an initial quench, the complex spin observable $F_\perp=F_x+iF_y$ is analyzed. \textbf{b) Scale separation} during the far from equilibrium evolution, encoded in the dual cascade of the particle distribution function $f(t,p)$. \textbf{c) Entropy transport} from low momenta (IR) to higher momenta (UV) with conserved total entropy $S$. For the systems considered, the total entropy can only be approximately estimated and the mutual entropy $S_{\mathrm{IR}\cap\mathrm{UV}}$ is small, such that $S\simeq S_\mathrm{IR}+S_\mathrm{UV}$. 
    \textbf{d) Entropy change} momentum profile  $\dot{\mathcal{S}}(t,p)$ drawn in the scaling regime. The vanishing integral of the curve (hatched areas) visualizes entropy conservation.}
    \label{fig:summary}
\end{figure*}
``For some reason, the universe at one time had a very low entropy for its energy content, and since then the entropy has increased." \cite{Feynman:2011} This paradigmatic statement of Richard Feynman reflects the observation that the world around us is so much structured rather than diffuse. The second law of thermodynamics states that the entropy of a closed system can only increase or remain constant; it never decreases. In the long run, by moving towards greater disorder or randomness, the entropic fate of the universe would then be inevitable. Following this reasoning, structures seen today may only exist because some low initial entropy deposit has not yet been used up~\cite{Penrose:1979}.

In this work we investigate an alternative scenario, which realizes that entropy can decrease on long-distance scales while it increases at short distances. While there is entropy transport from long to short distance scales, the total entropy remains constant: On a fundamental level, the universe evolves according to the laws of quantum physics, where the unitary time evolution leads to conservation of von Neumann entropy. The observation of macroscopically structured properties is then a consequence of a dynamical separation of scales rather than a low initial entropy deposit. Due to scale separation, the large-scale properties of the system become insensitive to the highly entropic microphysical processes, such that low-entropic macrophysics prevails.

To establish such a scenario for quantum many-body systems, we employ quantum simulations based on experiments with ultracold atoms, and calculations in quantum field theory from first principles. 
Since setups with ultracold atoms can be well isolated from the environment, they offer the possibility to study the entropy conserving dynamics of quantum systems. 
Together with well-controlled initial state preparation even far from equilibrium, and the ability to monitor the spatially resolved time evolution, these systems are very powerful in identifying the different contributions to entropy. 

On the theory side, we employ nonequilibrium effective action techniques based on a self-consistently resummed large-$N$ expansion to next-to-leading order in the number of field components $N$~\cite{Berges:2001fi}. This non-perturbative approach is entropy conserving, while it describes far-from-equilibrium dynamics and the approach to equilibrium. In contrast, simpler descriptions such as based on kinetic theory leading to Boltzmann equations are not entropy conserving in general, and may not be employed in this context.  

We investigate entropy transport for universal scaling phenomena far from equilibrium, which are abundant in nature. The universality has the advantage that the dynamics is insensitive to details of the governing Hamiltonian and nonequilibrium initial states. For a large class of different systems and initial conditions far from equilibrium, the same phenomenon is observed without any fine-tuning. 
For closed systems, universal far-from-equilibrium scaling associated to nonthermal fixed points were studied in detail theoretically~\cite{Berges:2008wm,Orioli:2015cpb,Chantesana:2018qsb,Berges:2020fwq} and experimentally~\cite{Prufer:2018hto,Erne:2018gmz,Glidden:2020qmu,Gazo:2023exc,Huh:2023xso}. A dynamical scale separation arises from the emergence of long-range order with condensation at zero momentum~\cite{Svistunov:1991,Bray:1994zz,Berloff:2002,Berges:2012us,Nowak:2012gd,Preis:2022uqs}, 
and energy transport towards high momenta due to wave turbulence~\cite{Nazarenko:2011,Micha:2004bv,Berges:2014bba}.

Experimentally, we consider a spinor Bose gas~\cite{Stamper-Kurn:2012bxy} of $^{87}$Rb atoms in a quasi-one-dimensional box-like trap~\cite {Prufer:2019kak}.
The dynamics is governed by three internal states, labeled by their magnetic quantum number $m \in {0,\pm1}$ as illustrated in Fig.~\ref{fig:summary}a. Initially, all atoms are prepared in the $m=0$ state, forming a spinor-condensate with zero spin length.  
By means of microwave dressing, a rapid change in the energy splitting of the states is induced such that spin excitations develop. Fig.~\ref{fig:summary}b sketches the dynamics of their distribution function $f(t,p)$ as a function of momentum $p$ for different times (see Methods for details). The quench at initial time induces excitations whose distribution is peaked at the coherence momentum scale $Q$, and a dynamical separation of scales 
emerges at later times.

We extract the entropy of the system using the Boltzmann-Einstein entropy in terms of the distribution function $f(t,p)$, and in addition Shannon's entropy based on a probability distribution estimated from the statistics of the experimental data. Remarkably, both estimates exhibit consistently the transport of entropy from low to high momentum scales, which is illustrated in Fig.~\ref{fig:summary}c. The shared information between the low- (IR) and high-momentum (UV) regimes turns out to be small for the self-similar scaling dynamics observed. Entropy conservation can then be directly observed from the cancellation of the negative and the positive contributions to the effective entropy change at low and high momenta, as sketched in Fig.~\ref{fig:summary}d.  

To indicate the scope of the experimental findings, and bridging to applications in the early universe, we theoretically analyze also a relativistic quantum field theory in $3+1$ space-time dimensions. We compute the non-equilibrium quantum dynamics for a scalar field theory as encountered in the Higgs sector of the Standard Model of particle physics or certain inflationary models. Once expressed in terms of a relativistic particle number distribution and entropy in three spatial dimensions, a similar entropy transport is observed as for the Bose gas schematically displayed in Fig.~\ref{fig:summary}b-d.

\section{Entropy}
\label{sec:entropy}

The von Neumann entropy $S$ of a quantum state described by the density operator $\hat\varrho$ is defined as
\begin{equation}\label{eq:vN_entropy}
    S=-\Tr\!\big(\hat\varrho\log\hat\varrho\big)\,.
\end{equation}
The dependence of the density operator on time $t$ is determined by the von Neumann equation 
\[i\partial_t \hat\varrho(t)=\Big[\hat H,\hat\varrho(t)\Big]\,.\]
Here the Hamilton operator $\hat H$ is time independent for a closed system. In thermal equilibrium the density operator is given in terms of the Hamiltonian, such that its time derivative vanishes. For nonequilibrium evolutions, the solution $\hat\varrho(t)$ at any time $t$ is determined by some initial state $\hat\varrho_0\equiv\hat\varrho(t_0)$ at a given starting time $t_0$. 
Most importantly for our purposes, the von Neumann entropy (\ref{eq:vN_entropy}) of any such solution $\hat\varrho(t)$ is constant in time,
\begin{equation}\label{eq:entropy_conservation}
    S=\mathrm{const}\,.
\end{equation}
This conservation of entropy reflects the fact that for the unitary time evolution of a closed quantum system there is no loss of information in any strict sense.

There are other notions of entropy, which do not employ the density operator describing the full quantum system. An example is the Boltzmann-Einstein entropy $H$, which employs a distribution function for the space-time and energy-momentum dependence of particles. For spatially homogeneous systems the time- and momentum-dependent distribution function\footnote{For inhomogeneous systems the distribution function can also depend on position.} 
\[f_{\bm{p}}\equiv f(t,{\bm{p}})\]
only depends on the modulus $p=|\bm{p}|$, with $f_{\bm{p}}=f_p$ if the system is also isotropic. 
For bosonic excitations the Boltzmann-Einstein entropy $H$ of the quantum many-body system is then defined as~\cite{Einstein:1925}
\begin{equation}\label{eq:EB_entropy}
    \begin{split}
    H&=\int_{\bm{p}}\Big((1+f_{\bm{p}})\log(1+f_{\bm{p}})-f_{\bm{p}}\log f_{\bm{p}}\Big)\,,
\end{split}
\end{equation}
which should not be confused with the Hamilton operator $\hat H$. Here we abbreviated the integration over momenta $\bm p$ in $d$ spatial dimensions as
\[\int_{\bm{p}}\equiv\int
\frac{\mathrm{d}^d\bm{p}}{(2\pi)^d}\,.\]
Introducing the mode entropy
\begin{equation} \label{eq:modeentropy}
    \mathcal{H}(f_{\bm{p}})=(1+f_{\bm{p}})\log(1+f_{\bm{p}})-f_{\bm{p}}\log f_{\bm{p}}\, ,
\end{equation}
we have 
\begin{equation} \label{eq:Hmode}
H=\int_{\bm{p}}\mathcal{H}(f_{\bm{p}})\propto\int_0^\infty\mathrm{d}p\,p^{d-1}\mathcal{H}(f_p)\,,
\end{equation}
where in the last relation, assuming isotropy, the angular integrals have been carried out. 

For the following, it will be instructive to compare also to Shannon's entropy~\cite{Shannon:1948}
\begin{equation}\label{eq:Shannon_entropy}
    I=-\sum_{\omega\in \Omega}P(\omega)\log P(\omega) \, ,
\end{equation}
which employs a probability distribution $P(\omega)$ 
for events $\omega$ drawn from a given sample space $\Omega$. This information theoretic entropy 
is closely related to the above quantum-statistical entropies.  

To make the comparison between the different notions of entropy more rigorous, we consider first the spinor Bose gas. Its Hamilton operator is described in terms of the spin operator $\hat{\bm{F}}$ with components $\hat{F}_x$, $\hat{F}_y$ and $\hat{F}_z$~\cite{Prufer:2018hto}. An observable, such as the two-point correlation function for the spin at time $t$, is given by the expectation value
\begin{equation} \label{eq:twopointcorrelator}
    \langle \hat{\bm{F}}^\dagger \hat{\bm{F}} \rangle  (t,\bm{x}-\bm{y}) = \Tr\Big( \hat{\varrho}(t) \hat{\bm{F}}^\dagger(\bm{x})\hat{\bm{F}}(\bm{y})\Big)\,. 
\end{equation}
More precisely, for the equal-time correlator we consider the symmetrized two-point function. 
The corresponding distribution function in momentum space is then obtained from its Fourier transform 
\begin{equation} \label{eq:distributionfunction}
    f(t,\bm{p}) + \frac{1}{2} = \int \mathrm{d}^d \bm{x}\, e^{-i\bm{p}\bm{x}}\langle \hat{\bm{F}}^\dagger \hat{\bm{F}} \rangle  (t,\bm{x}) \, , 
\end{equation}
where $f(t,\bm{p})$ takes into account the occupancies above the ground-state or vacuum ``quantum-half". 

In general, for an interacting system non-trivial higher correlation functions involving more than two spin operators, or other degrees of freedom not captured by the spin operators, exist. In this case the distribution function (\ref{eq:distributionfunction}), carrying only the information about the specific two-point correlation, does not provide a complete description. Correspondingly, the von Neumann entropy in terms of the full density operator (\ref{eq:vN_entropy}) and the Boltzmann-Einstein entropy including only the distribution function (\ref{eq:EB_entropy}) are not the same, $S\neq H$, in general.
In particular, rather than being conserved on a fundamental level as in (\ref{eq:entropy_conservation}), the Boltzmann-Einstein entropy can increase with time, known as the $H$-theorem~\cite{Boltzmann:1872,Tolman:1938}
\begin{equation}\label{eq:H_theorem}
    \frac{\mathrm{d}H}{\mathrm{d}t} \geq 0 \, .
\end{equation}
The equality in (\ref{eq:H_theorem}) holds only for $f_{p}=\mathrm{const}$.

However, there are special cases where von Neumann and Boltzmann-Einstein entropies do agree. The simplest examples are non-interacting systems, but also non-trivial mean-field or large-$N$ descriptions can exhibit this property. An example is given by the $N$-component field theory of Sec.~\ref{sec:simulation} if the limit $N \rightarrow \infty$ is considered.
In these cases the dynamics is Gaussian, which means that higher correlation functions factorize into lower ones.  For the special cases of Gaussian dynamics it can be shown that~\cite{Berges:2017hne} 
\begin{equation}\label{eq:Gaussianentropies}
S^{\text{(Gauss)}}=H^{\text{(Gauss)}} \, .
\end{equation}
Conservation of von Neumann entropy then implies that the $H$-theorem realizes the lower bound in (\ref{eq:H_theorem}), which entails a constant distribution function for Gaussian dynamics.

For interacting systems that are not described by 
Gaussian dynamics the equality (\ref{eq:Gaussianentropies}) no longer holds. Rather, it is convenient to write 
\begin{equation}\label{eq:BeyondGaussianentropies}
S = H + \Delta H \, ,
\end{equation}
where the correction $\Delta H$ comes from non-linear, beyond Gaussian contributions due to interactions. In a perturbative setting the Boltzmann-Einstein entropy $H$ may thus be viewed as the leading term in a series expansion of the von Neumann entropy $S$ around the Gaussian approximation if $\Delta H/H\ll 1$. From this perspective, the non-conservation or growth of Boltzmann-Einstein entropy as described by the $H$-theorem is a consequence of approximations. 

In turn, for a time evolving system out of equilibrium, the emergence of an approximately constant Boltzmann-Einstein entropy can indicate the approach to Gaussian dynamics. This is expected to be the case for the approach to the nonthermal scaling solutions of the spinor Bose gas and the relativistic field theory we consider. While these systems are strongly correlated, they dynamically approach an infrared fixed point for which Gaussian contributions can dominate.    

To discuss the information theoretic Shannon entropy $I$ for our system, we consider the experimental measurements for the spinor Bose gas based on positive operator valued measures (POVM)~\cite{Kunkel:2019vud}.
As illustrated in Fig.~\ref{fig:summary}a, the spatially resolved complex-valued transverse spin observable $F_\perp(t,x)= F_x(t,x) + i F_y(t,x)$ is measured. The square of the spin observable in momentum space,
$|F_\perp|^2(t,p)$, is then taken as the random variable for estimating Shannon's entropy (\ref{eq:Shannon_entropy}). This observable
takes on different values for each realization in the experiment. The probability distribution for all momenta is estimated from the statistics of the experimental data by considering a multivariate Gaussian distribution as discussed in the Methods. 

For the experimental data the average longitudinal spin may be neglected to good accuracy with $\langle |F_z|^2 \rangle \ll \langle |F_\perp|^2\rangle$. We will use this approximation in the following. Accordingly, we define in momentum space also the distribution function as the expectation value or experimental average $f(t,p)+1/2=\langle|F_\perp|^2\rangle(t,p)$ in view of~(\ref{eq:distributionfunction}). 
\vspace{-3mm}
\section{Scale dependent entropy}\label{sec:scaleentropy}
\vspace{-1mm}
In order to capture entropy transport across momentum scales, we define scale-dependent entropies for low momenta (IR) and for high momenta (UV) separated by a characteristic momentum scale $Q$. While we define this scale in more detail for different systems below, $Q$ turns out to be related to the inverse coherence (or healing) length scale for the spinor Bose gas~\cite{Stamper-Kurn:2012bxy}.

Taking Shannon's entropy as an example, any separation of entropy in infrared ($I_{\mathrm{IR}}$) and ultraviolet ($I_{\mathrm{UV}}$) components can be written in general as 
\begin{equation}\label{eq:mutualinformation}
I= I_{\mathrm{IR}} + I_{\mathrm{UV}} - I_{\mathrm{IR}\cap\mathrm{UV}} \, .
\end{equation}
The mutual information $I_{\mathrm{IR}\cap\mathrm{UV}}$ quantifies the information shared by the low- and high-momentum regimes. 
From the experimental data, we extract a probability distribution $P_\mathrm{IR}$ that yields the infrared entropy  
\begin{equation}
    I_\mathrm{IR}=-\sum_{\omega_\mathrm{IR}\in \Omega_\mathrm{IR}}P_\mathrm{IR}(\omega_\mathrm{IR})\log P_\mathrm{IR}(\omega_\mathrm{IR}) \,,
\end{equation}
and equivalently for the ultraviolet entropy $I_\mathrm{UV}$. 
We take the same spin observable in momentum space as in section~\ref{sec:entropy}, $|F_\perp(t,p)|^2$, but restricted to momenta $p \leq Q$ as the infrared  variable to obtain $P_\mathrm{IR}$. To extract $P_\mathrm{UV}$ we evaluate the spin observable for $p > Q$. The mutual information $I_{\mathrm{IR}\cap\mathrm{UV}}$ is then encoded in the IR and UV entropies together with the total entropy (\ref{eq:Shannon_entropy}), as explained in more detail in the Methods.

\begin{figure}[b]
    \includegraphics[width=.5\textwidth]{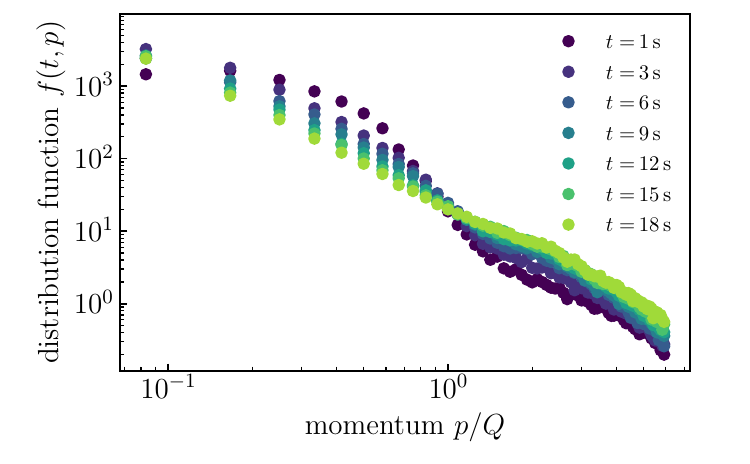}
    \caption{\textbf{Distribution function} $f(t,p)$ for the spinor Bose gas as a function of momentum for different times in the scaling regime.}
    \label{fig:f_Exp}
\end{figure}

For the Boltzmann-Einstein entropy (\ref{eq:EB_entropy}) we correspondingly partition the phase space of the system into a low- and high-momentum regime, separated by the characteristic  scale $Q$. The Boltzmann-Einstein entropy contained in the infrared regime is then given by
\begin{equation}\label{eq:IR_entropy}
    \begin{split}
    H_\mathrm{IR}&=\int_{|\bm{p}|\leq Q}\mathcal{H}(f_{\bm{p}})\propto\int_0^Q\mathrm{d}p\,p^{d-1}\mathcal{H}(f_p) \, .
    \end{split}
\end{equation}
Likewise, $H_\mathrm{UV}$ is determined from the integral including all momenta above the scale $Q$. Since
\begin{equation}
    H=H_\mathrm{IR}+H_\mathrm{UV} \, ,
\end{equation}
one immediately concludes that 
\begin{equation}
H_{\mathrm{IR}\cap\mathrm{UV}} = 0 \, ,
\end{equation}
i.e.~the mutual component vanishes identically for the Boltzmann-Einstein entropy.
Following the discussion of Sec.~\ref{sec:entropy} leading to (\ref{eq:Gaussianentropies}), it follows that for Gaussian dynamics the mutual component in momentum scale decompositions vanishes in general. In particular, we will demonstrate in the following that near the nonthermal fixed point of the spinor Bose gas the mutual information turns out to be relatively small, such that the Boltzmann-Einstein entropy can well approximate the von Neumann entropy.

\begin{figure*}
    \includegraphics[width=\textwidth]{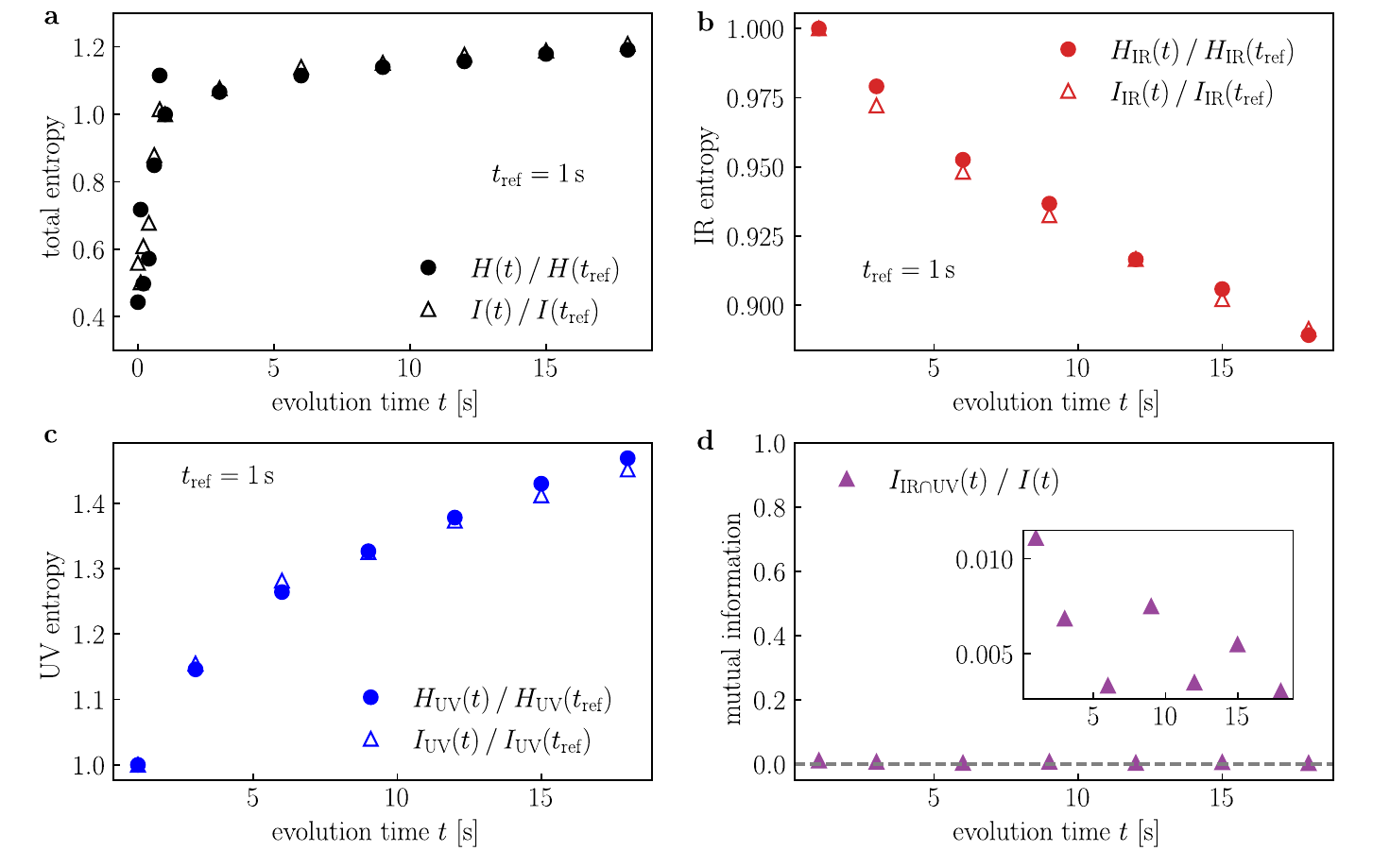}
    \caption{\textbf{a) Boltzmann-Einstein ($H$) vs.\ Shannon ($I$) entropy} estimates from the measured data as a function of time. An initial fast growth is followed by a relatively flat evolution after the time $t_\mathrm{ref}=1\,\mathrm{s}$, where the system enters the scaling regime.
    \textbf{b)~IR~entropies} are decreasing in the scaling regime. Both estimates for the IR Boltzmann-Einstein and Shannon entropies agree remarkably well, as expected even in this non-perturbative regime when the dynamics becomes Gaussian at low momenta. \textbf{c) UV entropies} increase correspondingly. Both estimates agree rather well in this perturbative regime. \textbf{d)~Relative mutual information} between the low- and high-momentum regimes. The inset displays the same data points on a sub-percentage level, showing that $I_{\mathrm{IR}\cap\mathrm{UV}}$ makes up only a small fraction of the total $I$.}
    \label{fig:experiment_summary}
\end{figure*}
\vspace{-2mm}
\section{Entropy transport: Experiment}
\label{sec:ExperimentalData}
\vspace{-2mm}
We consider the experimental setup for a spinor Bose gas in a quasi-one-dimensional trap of Ref.~\cite{Prufer:2019kak}, using the same measurement data. The system features ferromagnetic spin-spin interactions together with density-density interactions. 
An applied magnetic field induces a quadratic Zeeman shift, which breaks isotropy. The magnetic field is adjusted such that easy-plane ferromagnetic properties are observed. Our initial conditions restrict the dynamics to the spatially
averaged longitudinal ($z$-) spin being zero. 

A central quantity for our estimates of Boltzmann-Einstein's entropy $H$ is the distribution function defined from the complex spin variable, $f(t,p)=\langle|F_\perp|^2\rangle(t,p)-1/2$, as described in Sec.~\ref{sec:entropy}. 
Fig.~\ref{fig:f_Exp} shows the nonequilibrium evolution of the distribution function extracted from the experimental data for times between $1$\,s and $18$\,s. The momentum axis is normalized to the characteristic momentum scale $Q$ corresponding to the inverse healing length of the system~\cite{Prufer:2019kak}. One observes an approximately constant pivot point around $p/Q\simeq 1$, which separates a low-momentum scaling regime characterized by an inverse particle cascade from an energy cascade towards higher momenta as analyzed, e.g., in Ref.~\cite{Orioli:2015cpb}.\\
\begin{figure*}
    \includegraphics[width=\textwidth]{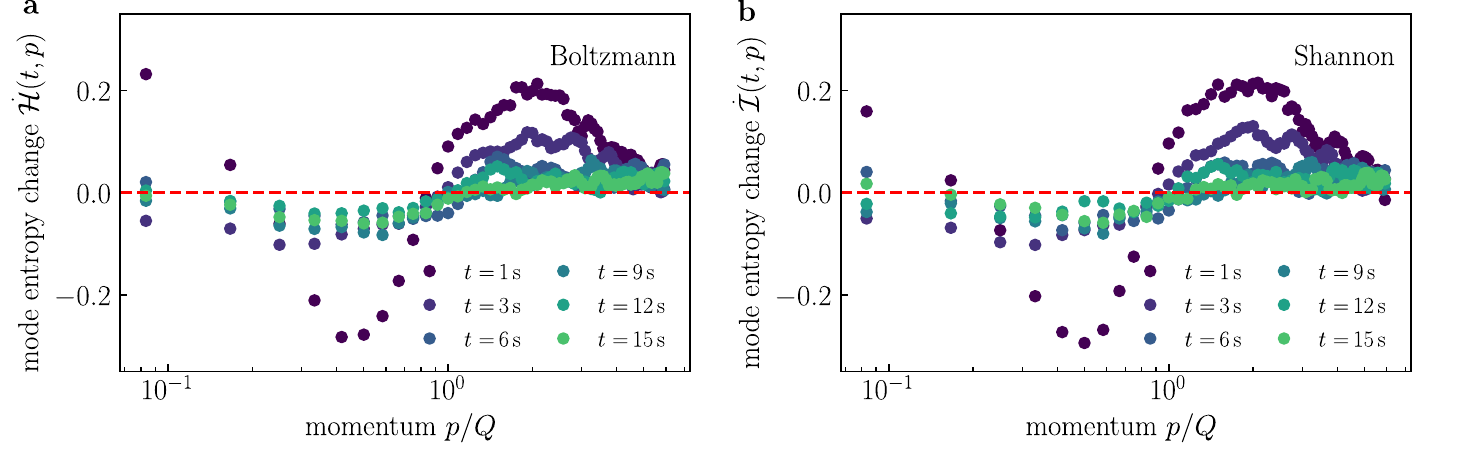}
    \caption{\textbf{a) Mode entropy change for the Boltzmann-Einstein entropy}, and \textbf{b) for the Shannon entropy} at different times in the scaling regime. Both estimates from the experimental data show great similarity, with the negative contributions occurring below the characteristic scale $Q$.}
    \label{fig:exp_entropy_density_change}
\end{figure*}
Following Sec.~\ref{sec:scaleentropy}, the scale $Q$ is used to define an IR entropy from low momenta and UV entropy from higher momenta for both Boltzmann-Einstein and Shannon entropies. In Fig.~\ref{fig:experiment_summary}a the time evolution of the (total) Boltzmann-Einstein and Shannon entropies of the system are shown. We normalize to the respective values at $t_\mathrm{ref}=1\,\mathrm{s}$ where the system starts to exhibit scaling. One observes that the entropies rapidly increase for early times. This initial increase indicates an incomplete description, including also sizable non-Gaussianities at early times, which leads to the expected large deviations from the conserved von Neumann entropy. When the system evolves in the scaling regime after $t_\mathrm{ref}=1\,\mathrm{s}$, both entropies exhibit significantly less variation in time. 

This observation is in accordance with the approach to an infrared fixed point for which Gaussian contributions are expected to dominate. Following Sec.~\ref{sec:entropy}, in this case Boltzmann-Einstein and Shannon entropies in the infrared should agree to good accuracy. Fig.~\ref{fig:experiment_summary}b shows the IR entropies, which indeed turn out to agree remarkably well with deviations of less than a percent. The data clearly demonstrates the decrease of entropy in the low-momentum regime, respectively on long-distance scales.

As shown in Fig.~\ref{fig:experiment_summary}c, 
the agreement between Boltzmann-Einstein and Shannon entropies is also remarkably good for the UV entropy, where the perturbative wave-turbulent energy cascade occurs. 
The UV entropies exhibit a monotonic increase over the whole time interval.

We also extract the mutual information from the data by considering the statistics of all momentum modes as described in Sec.~\ref{sec:methods}. 
From Fig.~\ref{fig:experiment_summary}d one observes that in the scaling regime the relative mutual information becomes smaller than one percent,  
suggesting a separate consideration of the low- and high-momentum regime. 

To get a refined picture, we analyze the entropy transport mode by mode in momentum space. For this we consider the temporal change of mode entropy, which for the Boltzmann-Einstein entropy is given by the time derivative of (\ref{eq:modeentropy}). 
Fig.~\ref{fig:exp_entropy_density_change}a shows the momentum profile of the Boltzmann-Einstein entropy change for different times in the scaling regime. We compare the result with the corresponding Shannon entropy change $\dot{\mathcal{I}}(t,p)$ as introduced in Sec.~\ref{sec:methods}. In both cases we have performed an averaging of the entropy change over neighboring momenta to smooth the experimental data. One observes a very good qualitative agreement between the different mode entropies. Above the characteristic scale $Q$ the change is positive for all times. Below the scale $Q$, however, the change becomes negative with the exception of momentum modes in the deep infrared, in particular at earlier times. The negative contributions dominate in the low-momentum regime, leading to the IR entropy decrease seen in Fig.~\ref{fig:experiment_summary}b. Correspondingly, the positive contributions above the scale $Q$ lead to the increasing UV entropy encountered in Fig.~\ref{fig:experiment_summary}c.
\vspace{-2mm}
\section{Entropy transport in relativistic quantum field theory}\label{sec:simulation}
%\vspace{-2mm}
\begin{figure}[b]
    \includegraphics[width=1.02\linewidth]{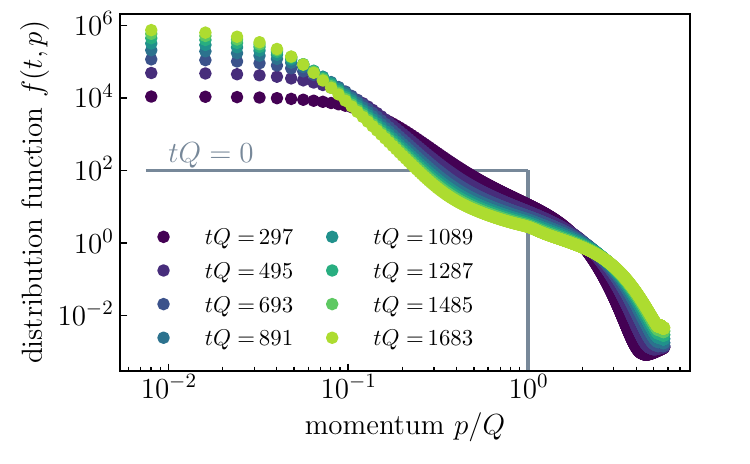}
    \caption{\textbf{Distribution function} as a function of momentum at different times for the relativistic quantum field theory. The time evolution exhibits a dynamical separation of scales, with transport of particles towards low momenta and energy transport towards high momenta. The initial distribution with momentum scale $Q$ is given by the gray line.}
    \label{fig:f_2PI}
\end{figure}
In the following, we investigate entropy transport for relativistic scalar fields in $3+1$ space-time dimensions. The theory is $O(N)$ symmetric for a field operator $\hat{\phi}_a(t,\bm{x})$ with $a = 1,\ldots,N$ components and quartic self-interaction. We employ a self-consistent expansion to next-to-leading order in $1/N$~\cite{Berges:2001fi}. For the results, we specify to $N=4$ as encountered in the Higgs sector of the Standard Model of particle physics or certain inflationary models for early universe dynamics~\cite{berges2015}.

\begin{figure*}[t]
    \includegraphics[width=0.98\textwidth]{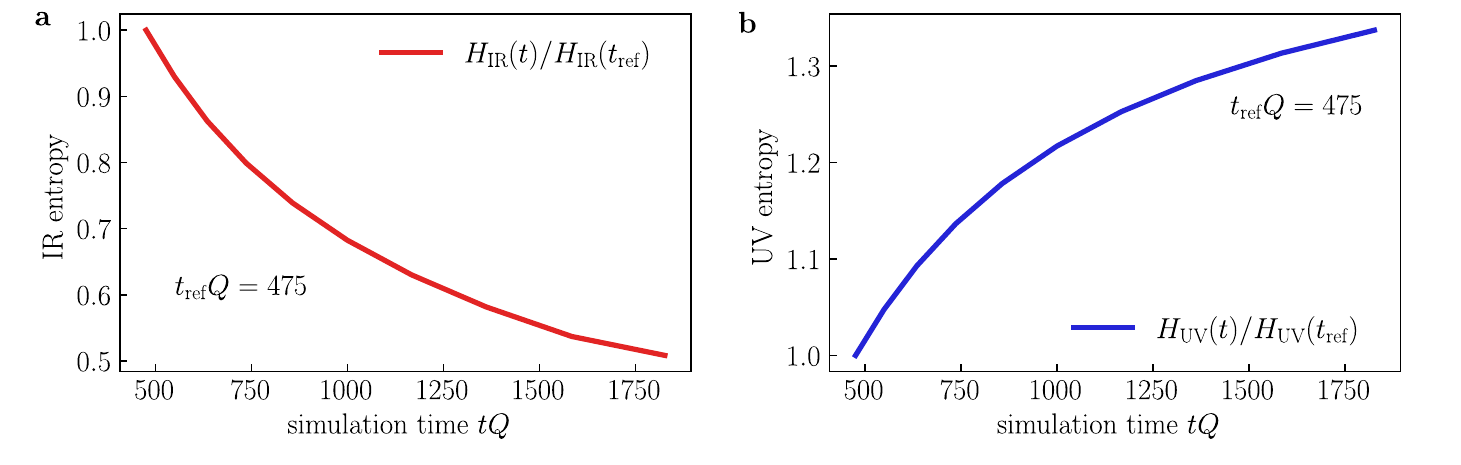}
    \caption{\textbf{a) IR entropy} in $3+1$ dimensions as a function of time. Similar to the non-relativistic Bose gas, the IR entropy of the relativistic field theory monotonically decreases while the \textbf{b)~UV entropy} is monotonically increasing with time.} 
    \label{fig:2PI_entropies}
\end{figure*}
The distribution function $f(t,\bm p)$ is obtained from the spatial Fourier transform of the (symmetrized) two-point correlation function $\langle \hat{\phi}_a(t,\bm{x}) \hat{\phi}_a(t,\bm{y})\rangle$ and time derivatives as  
described in Sec.~\ref{sec:methods}.
The evolution of the distribution function as a function of momentum $p = |\bm p|$ at different times is shown in Fig.~\ref{fig:f_2PI}, starting far from equilibrium with high occupancies up to the momentum scale $Q$. Very similar to the non-relativistic Bose gas data of Sec.~\ref{sec:ExperimentalData}, a dynamical separation of scales emerges: The low-momentum regime exhibits an increasingly high occupation of modes due to the transport of particles towards lower momenta. In turn, at higher momenta an energy cascade towards the UV occurs. 

The two momentum regimes are separated by an intermediate momentum scale $Q(t)$, which can dependent on time. We extract this scale as described in the Methods. At initial time $Q(t=0) =Q$ characterizing the box-like far-from-equilibrium initial condition for the distribution function shown in Fig.~\ref{fig:f_2PI}. We find with $Q(t) \simeq \mathrm{const}$ that the intermediate momentum scale is approximately constant in time, 
moving slightly towards lower momenta during the nonequilibrium evolution. 
\begin{figure*}
    \includegraphics[width=\textwidth]{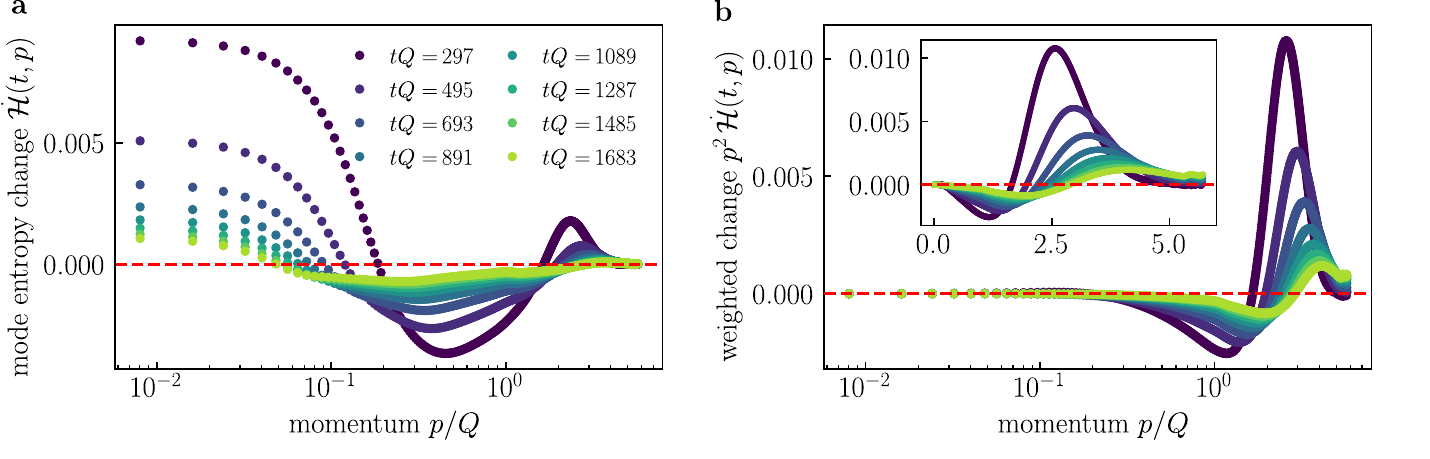}
    \caption{\textbf{a) Mode entropy change} $\dot{\mathcal{H}}(t,p)$ as a function of momentum at different times, showing positive contributions in the high- as well as low-momentum regimes, and negative contributions in between. \textbf{b) Weighted mode entropy change} including the phase-space factor $\sim p^2$ in three spatial dimensions. The inset shows the same plot on a linear momentum axis. In the scaling regime at later times the areas below and above the momentum axis and the curves become equal, illustrating conservation of Boltzmann-Einstein entropy.}
    \label{fig:2PI_entropy_change}
\end{figure*} 

In the following, we extract the Boltzmann-Einstein entropy $H$ for the relativistic field theory. 
Fig.~\ref{fig:2PI_entropies}a shows the entropy contained in the low-momentum regime as defined in~(\ref{eq:IR_entropy}).
Similar to the results for the Bose gas before, one observes a monotonic decrease of entropy in the low-momentum regime. 
In Fig.~\ref{fig:2PI_entropies}b we consider the entropy contained in the high-momentum regime of the system. Similar to the non-relativistic gas in Sec.~\ref{sec:ExperimentalData}, in Fig.~\ref{fig:experiment_summary}c one observes a monotonic increase also for the relativistic system.

A more detailed picture is obtained by analyzing the entropy transport mode by mode in momentum space. For this we consider
in Fig.~\ref{fig:2PI_entropy_change}a the time derivative $\dot{\mathcal{H}}(t,p)$ of the mode entropy 
introduced in~(\ref{eq:modeentropy}). Similar to the experimental data of the Bose gas shown in Fig.~\ref{fig:exp_entropy_density_change}, we observe for the relativistic field theory positive changes of the mode entropy at high momenta 
as well as in the deep infrared, while there are negative changes in between. 

In Fig.~\ref{fig:2PI_entropy_change}b we show the same quantity but rescaled as $p^2 \dot{\mathcal{H}}(t,p)$, i.e.~including the phase-space factor $\sim p^2$ coming from the measure of the momentum integral in~(\ref{eq:Hmode}) for the theory in $d=3$ spatial dimensions. This is different for the experimental Bose gas setup considered in Sec.~\ref{sec:ExperimentalData}, which is effectively one-dimensional. One observes that the positive contributions to the entropy change at sufficiently low momenta are effectively suppressed because of phase space in the three-dimensional case. This results in a relatively strong decrease of 
the IR entropy in $3+1$ dimensions as compared to the Bose gas case. 
  
Since the IR part of the entropy decreases and the UV part increases, there is a continuous transport of entropy towards higher momenta. This leads to a depletion at low momenta and the accumulation of entropy at higher momenta or short-distance scales, respectively. More specifically, Fig.~\ref{fig:2PI_entropy_change}b shows that the peak values of the weighted entropy change $p^2 \dot{\mathcal{H}}(t,p)$ move to larger momenta with time, such that the dominant changes effectively occur above the scale $Q$.
The inset of Fig.~\ref{fig:2PI_entropy_change}b shows the same as the main plot, but on a linear rather than logarithmic momentum axis. The linear plot illustrates the approximate conservation of Boltzmann-Einstein entropy with the approach of the system to the scaling regime: At early times the positive contributions of the integral at high momenta clearly dominate over the negative contributions at low momenta, leading to an initial increase of Boltzmann-Einstein entropy $H$ similar to the case of the Bose gas discussed before. In contrast, at later times one observes from Fig.~\ref{fig:2PI_entropy_change}b that the positive and negative contributions to the integral begin to become equal in size with opposite sign, expressing the conservation of Boltzmann-Einstein entropy. This occurs once the approximately Gaussian scaling solution is approached in the infrared, such that the Boltzmann-Einstein entropy can provide a good estimate for the conserved von Neumann entropy.
\vspace{-2mm}
\section{Discussion and outlook}\label{sec:outlook}
%\vspace{-2mm}
Our analysis demonstrates that entropy can decrease on long-distance scales for universal phenomena far from equilibrium. Scaling phenomena like those we consider are abundant in nature, and our results are directly applicable to a wide scope of systems in the same universality class. These range from highest energies in early-universe cosmology or relativistic collisions in particle physics, to lowest energies in ultracold quantum gases. Accordingly, we exemplify our findings from experimental data of a spinor Bose gas, and from ab initio calculations in a relativistic quantum field theory. 

These examples exhibit a dynamical separation of scales far from equilibrium. The conservation of von Neumann entropy then directly implies a reciprocal connection between entropy decrease on long-distance scales and increase at short distances: One of them cannot exist without the other due to a conservation law on a fundamental level. Therefore, the observation of disorder on smaller distance scales {\it predicts} also emerging long-range order in these cases, and vice versa. 

For the considered scaling phenomena we demonstrate the connection between the emergence of macroscopic long-range order and microscopic disorder. This becomes possible since for scaling phenomena described by (near) Gaussian fixed points, the von Neumann entropy (approximately) equals the Boltzmann-Einstein entropy. The latter is accessible from spatially resolved measurements, and also ab initio computations, of low-order correlation functions which we employ. 

Accordingly, we uncover that the mutual information between the low- and high-momentum regimes is small. 
In particular, the large-scale properties become insensitive to the highly entropic microphysical processes, such that low-entropic macrophysics prevails in effective descriptions. The notion of effective  theories for many-body systems is central for the understanding of emergent phenomena in physics. Our analysis exemplifies how low-entropy effective descriptions can arise dynamically without the need for a specific initial low-entropy deposit.   

We emphasize that our results cannot be obtained from conventional descriptions, such as based on standard kinetic theory or the Boltzmann equation.
In particular, the observations violate the ``strong $H$-theorem", stating that the mode entropy 
(\ref{eq:modeentropy}) could never decrease.
The latter statement is indeed fulfilled for a large class of kinetic theories as shown in the Methods, which includes Boltzmann's equation for two-to-two scatterings. In contrast, the measurements for the spinor Bose gas and the quantum field theoretical calculations for the relativistic system provide explicit examples that low-momentum contributions to entropy can decrease. This is crucial to realize total entropy conservation in closed systems during nonequilibrium evolutions. 

Moreover, small deviations from thermal equilibrium, such as described by linear response theory, in general do not exhibit a dynamical separation of scales. Linear response around thermal equilibrium is fully determined in terms of thermal correlation functions, which are themselves time-translation invariant. Accordingly, our analysis in this work concerns genuine nonequilibrium phenomena that cannot be obtained from small perturbations of thermally equilibrated systems. For a wide class of far-from-equilibrium initial conditions, the subsequent non-linear dynamics of the quantum many-body systems is not characterized by relaxation as, e.g., would be the case in linear response. 

It is an exciting question how long these remarkable nonequilibrium phenomena last in time. Is the emergence of the low-entropic effective description a transient phenomenon, or could the ``critical slowing-down" of scaling phenomena delay its ending? 
For the universal scaling phenomena the time $t_*$ for the build-up of long-range order scales with the volume $V$ as $t_* \sim V^{1/\alpha}$ with universal scaling exponent $\alpha >0$, which is given by $\alpha = d/2$ for our systems in $d$ spatial dimensions~\cite{Orioli:2015cpb}. The long-distance scaling phenomenon has to end once the entire volume becomes correlated. In turn, for sufficiently large volumes, or even the entire universe as an infinite system, in practice it may not end for times of interest. 

Stated more broadly, an answer to the question about the entropic fate of the universe as a (quantum) many-body system has to take into account possible emergent phenomena. Entropic estimates based only on growth arising from the microphysical degrees of freedom can be misleading. Collective behavior can give rise to low-entropic macrophysics, whose effective physical laws can lead to new -- and sometimes even diverging -- time scales. 

While it is remarkable that some of these questions can be studied in detail in quantum systems from first principles, as exemplified in this work, very similar observations hold also in classical systems with many degrees of freedom and scale separations. An example is given for a corresponding classical field theory in the Methods section. This opens up a very wide range of applications of our findings, towards an answer to the question of why the world around us is so much structured rather than diffuse.

\acknowledgements

We thank M.~Heinrich, J.~Gebhard, Y.~Deller and S.~Lannig for discussions.
This work is part of
and funded by the Deutsche Forschungsgemeinschaft (DFG,
German Research Foundation) through the Collaborative Research
Centre, Project-ID No.\ 273811115, SFB 1225 ISOQUANT, and 
Germany’s Excellence
Strategy EXC 2181/1–390900948 (the Heidelberg STRUCTURES
Excellence Cluster).
The authors acknowledge support by the state of Baden-Württemberg through bwHPC.

\section{Methods}\label{sec:methods}
\begin{figure*}
    %\includesvg[width=\textwidth]{figures/CS/CS_entropy_change}
    \includegraphics[width=\textwidth]{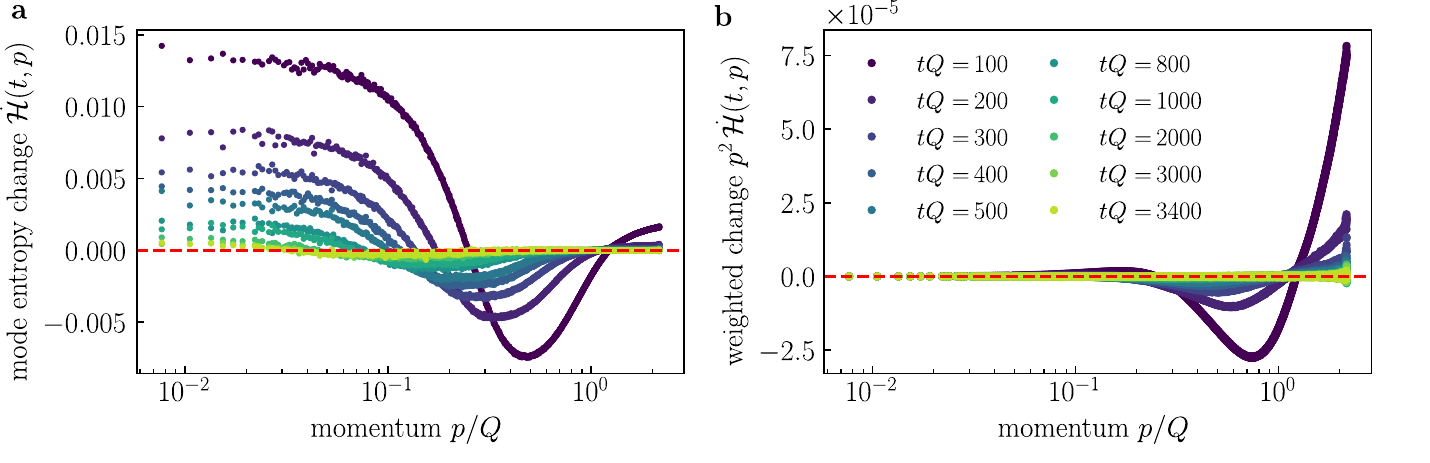}
    \caption{\textbf{a) Mode entropy change} $\dot{\mathcal{H}}(t,p)$ for a self-interacting classical-statistical field theory in $3+1$ space-time dimensions. \textbf{b) Weighted mode entropy change} including the phase-space factor $\sim p^2$. Below the characteristic momentum scale $Q$, one observes the behavior as for the corresponding quantum theory shown in Fig.~\ref{fig:2PI_entropy_change}a. However, in the high momentum regime, rather than the peak seen both in the quantum theory and the experiment, for the classical-statistical theory one observes a Rayleigh-Jeans-like behavior.} 
    \label{fig:CS_entropy_change}
\end{figure*}

\subsection{Shannon entropy estimation}
In order to estimate the Shannon entropy $I$ defined in Eq.~(\ref{eq:Shannon_entropy}), we consider the observables $|F_\perp|^2(p_j)$ at given time $t$. For each of the discrete momentum modes $p_j$ with $j=1,\ldots,M$ the observable $|F_\perp|^2(p_j)$ takes on different values for each realization of the experimental measurement in general.  
Assuming that the measured values for $|F_\perp|^2(p_j)$ are normally distributed, for the total system including all momentum modes $p_j$ we make the ansatz of a multivariate normal distribution 
\begin{equation}\label{eq:GaussianDistribution}
    P(\omega)=\frac{1}{\mathcal{N}}\exp(-\frac{1}{2}(\omega-\mu)^\mathrm{t}\,\Sigma^{-1}(\omega-\mu))
\end{equation}
with normalization $\mathcal{N}$, mean $\mu$ and covariance matrix $\Sigma$. Here the $M$-component events $\omega$ of the sample space $\Omega$ are given by
\[\omega=\begin{pmatrix}|F_\perp|^2(p_{\min})\\
\vdots\\|F_\perp|^2(p_{\max})\end{pmatrix}.
\]
For a multivariate normal distribution, the Shannon entropy is completely determined by the covariance matrix\footnote{We estimate the covariance matrix from the experimental data using the \texttt{Python}-package \texttt{numpy.cov()}.} $\Sigma$ and is given by~\cite{30996}
\[I=\frac{1}{2}\log((2\pi e)^M\!\det(\Sigma))\,.\]

Since different probability distributions might give rise to the same measurement values of an observable, due to the finite statistics of the experimental data there is some ambiguity in how one estimates the Shannon entropy of a given data set. Aside from the Gaussian ansatz presented above we considered as a check also a non-parametric (Kozachenko–Leonenko) entropy estimator \cite{versteeg2014}. We find very similar results with both approaches, specifically for the mode entropy change in Fig.~\ref{fig:exp_entropy_density_change}b. The non-parametric entropy estimator has also been used for an entropy study of the spinor Bose gas considered in this work in position space in Ref.~\cite{Deller_2025}. In that entropy estimator the probability density $P(\omega)$ of continuous distributions is estimated from the distance between $\omega$ and its $k^\mathrm{th}$ nearest neighbors in the data set (with, e.g., $k=3$) as described in Ref.~\cite{kozachenko1987}. 

We split the sample space $\Omega$ into low- and high-energy subspaces, which are separated by the intermediate scale $Q$, by distinguishing $\omega_\mathrm{IR}=\begin{pmatrix}|F_\perp|^2(p_{\min})&\cdots&|F_\perp|^2(Q)\end{pmatrix}^\mathrm{t}$ from the corresponding $\omega_\mathrm{UV}$. With this partition we have
\[\omega=\begin{pmatrix}\omega_\mathrm{IR}\\\omega_\mathrm{UV}\end{pmatrix},\]
corresponding to $\Omega=\Omega_\mathrm{IR}\cup\Omega_\mathrm{UV}$. 
The mutual information in~(\ref{eq:mutualinformation}) can then be computed as
\[\begin{split}
    I_{\mathrm{IR}\cap\mathrm{UV}}&=\sum_{\omega\in\Omega}P(\omega)\log(\frac{P(\omega)}{P_\mathrm{IR}(\omega_\mathrm{IR})P_\mathrm{UV}(\omega_\mathrm{UV})})\,.
\end{split}\]
Note that it vanishes when the joint probability distribution $P(\omega)$ factorizes into the marginal distributions $P_\mathrm{IR}(\omega_\mathrm{IR})$ and $P_\mathrm{UV}(\omega_\mathrm{UV})$,
\[I_{\mathrm{IR}\cap\mathrm{UV}}=0\quad\text{for}\quad P(\omega)=P_\mathrm{IR}(\omega_\mathrm{IR})P_\mathrm{UV}(\omega_\mathrm{UV})\,.\] 

Finally, the mode entropy $\mathcal{I}(p)$ considered in Fig.~\ref{fig:exp_entropy_density_change}b is estimated from the statistics of a single momentum-mode observable $|F_\perp|^2(p)$.

\subsection{Relativistic field theory in $3+1$ dimensions}
We consider a relativistic, $N$-component real scalar field theory in $3+1$ space-time dimensions. The $O(N)$-symmetric Hamiltonian is given by
\begin{equation}\label{eq:HamiltonianO(N)}
    \begin{split}
        \hat H&=\int_{\bm x}\bigg\{\frac{1}{2}\hat\pi_a(\bm x)\hat\pi_a(\bm x)+\frac{1}{2}\nabla\hat\phi_a(\bm x)\nabla\hat\phi_a(\bm x)\\
        &\qquad+\frac{m^2}{2} \hat\phi_a(\bm x)\hat\phi_a(\bm x)+\frac{\lambda}{4!N}\big(\hat\phi_a(\bm x)\hat\phi_a(\bm x)\big)^2\bigg\} \, ,
    \end{split}
\end{equation}
where repeated indices $a=1,\cdots,N$ are summed over throughout.
The conjugate field operators $\hat\phi_a(\bm x)$ and $\hat\pi_a(\bm x)$ satisfy the canonical commutation relations 
\[\Big[\hat\phi_a(\bm x),\hat\pi_b(\bm y)\Big]=i\delta_{ab}\delta(\bm x-\bm y)\,.\]

For the Heisenberg field operators we write $\hat\phi_a(x)\equiv\hat\phi_a(x^0,\bm x)$.
The symmetrized and anti-symmetrized two-point correlations are given by \cite{berges2015}
\begin{equation}\label{eq:StectralAndStatisticalFunction}
    \begin{split}
    F_{ab}(x,y)&=\frac{1}{2}\Big\langle\Big\{\hat\phi_a(x),\hat\phi_b(y)\Big\}\Big\rangle\,,\\
    \rho_{ab}(x,y)&= i\,\Big\langle\Big[\hat\phi_a(x),\hat\phi_b(y)\Big]\Big\rangle\,,
\end{split}
\end{equation}
using $\langle\hat\phi_a(x)\rangle=0$.
Here we employ $\langle\cdot\rangle=\Tr(\hat\varrho_0\,\cdot)$ 
for a density operator $\hat\varrho_0\equiv\hat\varrho(t_0)$ specifying the initial state of the system. We consider spatially homogeneous systems, where
\[F_{ab}(x,y)=F(x^0,y^0;\bm x-\bm y)\,\delta_{ab}
\]
and similarly for the spectral function $\rho_{ab}(x,y)=\rho(x^0,y^0;\bm x-\bm y)\,\delta_{ab}$.
The distribution function $f(t,\bm p)$, encoding the (statistical) fluctuations in the system, is given by the Fourier modes of the equal-time statistical function,
\begin{equation*}
    F(t,t;\bm p)=\int_{\bm x}e^{-i\bm p\bm x}F(t,t;\bm x)\,,
\end{equation*}
as well as their time derivatives, 
\begin{equation}\label{eq:DistributionFunctionFieldTheory}
    f(t,\bm p)+\frac{1}{2}=\sqrt{F(t,t;\bm p)\ddot{F}(t,t;\bm p)- \dot{F}^2(t,t;\bm p)}\,.
\end{equation}
Here 
\begin{eqnarray}
    \ddot{F}(t,t;\bm p) &\equiv& \partial_t\partial_{t^\prime} F(t,t^\prime;\bm p)\Big|_{t^\prime=t} , \nonumber\\
    \dot{F}(t,t;\bm p) &\equiv& \frac{1}{2}\Big(\partial_t F(t,t^\prime;\bm p)+\partial_{t^\prime} F(t,t^\prime;\bm p)\Big)\Big|_{t^\prime=t}\, . \nonumber 
\end{eqnarray}
The statistical and spectral function in~(\ref{eq:StectralAndStatisticalFunction}) are found by solving the coupled evolution equations~\cite{berges2015}
\begin{equation}\label{eq:PropagatorEvolutionEquations}
    \begin{split}
    \big(\Box_x\delta_{ac}+M^2(x)\big)F_{cb}(x,y)&=-\int_{t_0}^{x^0}\!\mathrm{d}z\,\Sigma^{\rho}_{ac}(x,z)F_{cb}(z,y)\\
    &\quad+\int_{t_0}^{y^0}\!\mathrm{d}z\,\Sigma^F_{ac}(x,z)\rho_{cb}(z,y)\\
    \big(\Box_x\delta_{ac}+M^2(x)\big)\rho_{cb}(x,y)&=-\int_{x^0}^{y^0}\!\mathrm{d}z\,\Sigma^\rho_{ac}(x,z)\rho_{cb}(z,y)\,,
\end{split}
\end{equation}
with d'Alembertian $\Box_x=\frac{\partial}{\partial x_\mu}\frac{\partial}{\partial x^\mu}$, time-dependent mass square $M^2(x)\equiv m^2+\Sigma^{(0)}(x)$ and short-hand $\int_{t_1}^{t_2}\mathrm{d}z\equiv\int_{t_1}^{t_2}\mathrm{d}z^0\int_{\bm z}$. Here the local ($\Sigma^{(0)}$), statistical ($\Sigma^F$) and spectral ($\Sigma^\rho$) parts of the self-energy $\Sigma$, as introduced in Ref.~\cite{berges2015}, depend on the statistical and spectral functions, $\Sigma\equiv\Sigma[F,\rho]$, coupling the two equations.  

To solve the evolution equations~(\ref{eq:PropagatorEvolutionEquations}), we truncate the self-energy $\Sigma$ at next-to-leading order (NLO) in $1/N$ as described in Ref.~\cite{Berges:2001fi}. 
Using code developed in Ref.~\cite{Shen_2020}, the distribution function $f(t,\bm p)$ in Fig.~\ref{fig:f_2PI} is obtained for $N=4$ field components, quartic coupling $\lambda=0.01$ and mass parameter square $m^2=0$ for box-like initial conditions, with box height $A_0=1\times10^{4}$ and width $Q$.

\subsection{Comparison to classical-statistical field theory}
 
In the following, we consider a relativistic classical-statistical field theory in $3+1$ space-time dimensions. The classical field equation of motion,
\begin{equation}
    \Big(-\Box_x-m^2-\frac{\lambda}{6N}\varphi_b(x)\varphi_b(x)\Big)\varphi_a(x)=0\,,
\end{equation}
of the $O(N)$ symmetric theory is solved repeatedly for different initial conditions. The initial field configurations $\varphi_{a,0}\equiv\varphi_a(t_0)$ and corresponding time-derivative configurations $\dot\varphi_{a,0}\equiv\varphi_a(t_0)$ are sampled. 
The width of the initial Gaussian distribution is given by some initial distribution function $f(t_0,\bm p)$ as described in detail in Ref.~\cite{Orioli:2015cpb,berges2015}. Solving the equations of motion for each of $R$ samples of the initial conditions, giving rise to the $R$ solutions $\varphi_a(t)$, the equal-time statistical function is given by
\[F(t,t;\bm p)=\int_{\bm x}e^{-i\bm p\bm x}\frac{1}{N}\big\langle\varphi_a(t,\bm x)\varphi_a(t,\bm 0)\big\rangle\,.\]
Here angle brackets denote averaging over the $R$ realizations, $\langle\cdot\rangle=\frac{1}{R}\sum_{r=1}^R$. As for the relativistic quantum theory, we can extract the distribution function $f(t,\bm p)$ from the equal-time statistical function $F(t,t;\bm p)$ according to ~(\ref{eq:DistributionFunctionFieldTheory}). 

\begin{figure}
    \centering
    \includegraphics[width=\linewidth]{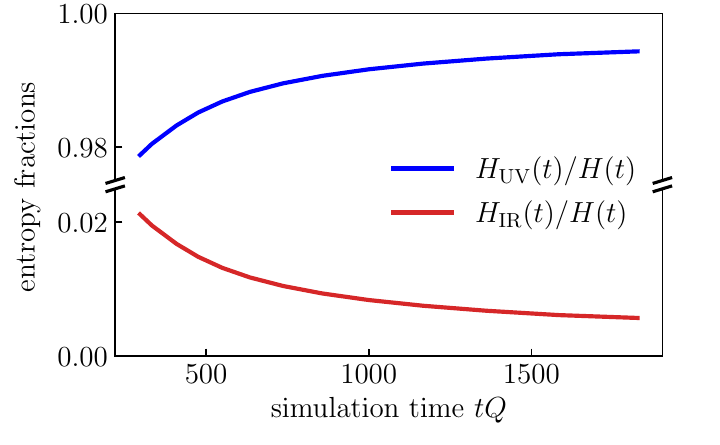}
    \caption{\textbf{IR and UV entropy fractions} of the total Einstein-Boltzmann entropy for the $3+1$ dimensional quantum field theory as a function of time. 
    From the decreasing fraction of the IR entropy to the total entropy of a few percent only, one observes that the entropy contributions are dominated by high momenta, i.e.\ small length scales.}
    \label{fig:entropy_fractions}
\end{figure}
Specifying to $N=4$ field components with quartic coupling $\lambda=0.1$ and mass parameter square $m^2=0.01$, in Fig.~\ref{fig:CS_entropy_change}a we show the mode entropy change $\dot{\mathcal{H}}(t,p)$ of the classical-statistical simulation for box-like initial conditions (of height $A_0=125$), obtained for $R=22$ realizations. The code used for obtaining the classical-statistical distribution function $f_p$ is taken from Ref.~\cite{Orioli:2015cpb}. In the low-energy regime, i.e.\ below the characteristic momentum $Q(t)$, one observes similar results as for the corresponding quantum theory in Fig.~\ref{fig:2PI_entropy_change}a. This is to be expected since in the low-momentum regime the occupation numbers are large, 
\begin{equation}\label{eq:ClassicalityCondition}
    f(t,\bm p)\gg1\,,
\end{equation}
leading to a classical-statistical behavior of the quantum system as discussed in Ref.~\cite{berges2015}. In contrast, in the high-energy regime of the system, i.e. above the characteristic momentum $Q(t)$, the clear peak observed both in the quantum theory (Fig.~\ref{fig:2PI_entropy_change}a) and experiment (Fig.~\ref{fig:exp_entropy_density_change}) is not resolved in the classical-statistical theory. As shown more prominently by the weighted mode entropy change $p^2\dot{\mathcal{H}}_p$ in Fig.~\ref{fig:CS_entropy_change}b we observe a Rayleigh-Jeans-type divergence of the classical-statistical theory in the high-energy regime. In the quantum system the unphysical divergence is cured, which becomes important at high energies since for sufficiently high momentum modes the classicality condition~(\ref{eq:ClassicalityCondition}) is no longer satisfied. 

\begin{figure}
    \includegraphics[width=\linewidth]{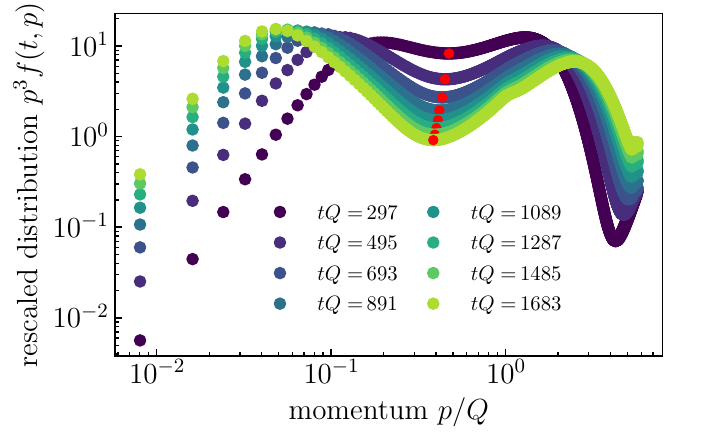}
    \caption{\textbf{Rescaled distribution function} $p^3f(t,p)$ as a function of momentum for different times. The local minima (red dots) of the rescaled distribution function are used to identify the intermediate scale $Q(t)$, separating the low and high momentum regimes. The scale $Q(t)$ varies slowly with time, slightly shifting towards lower momenta.} 
    \label{fig:f_resc_2PI}
\end{figure}
As shown in Fig.~\ref{fig:entropy_fractions} for the quantum theory in $3+1$ space-time dimensions, the predominant part of the system's entropy is contained in the high-momentum regime. With the entropy transport from low- to high-momentum modes the entropy fraction of the UV regime increases while the corresponding IR fraction decreases with time in accordance with Fig.~\ref{fig:2PI_entropies}. While classical-statistical simulations do not capture the high-entropy contributions in the high-momentum regime of the quantum system (dominating the total entropy), they can still be used to study ordering processes in the low-energy regime, as associated to a decreasing (mode) entropy. Although emerging large-scale ordering does in general not dominate the total entropy content, it can still dominate the effective macroscopic properties of the system. 

The scale $Q(t)$, separating the low- and high-energy regimes of the system, is for the relativistic field theories considered here identified by finding the local minimum of the rescaled distribution function $p^3 f_p$ as shown in Fig.~\ref{fig:f_resc_2PI}.

\subsection{H-theorem}
\label{app:H-theorem}
We show that in kinetic theory, as described by the Boltzmann equation \cite{Boltzmann:1872,uehling1933transport}
\begin{equation}\label{eq:BoltzmannEquation}
    \partial_t f_{\bm{p}}=C[f](\bm p) \, ,
\end{equation}
the mode entropy $\mathcal{H}_{\bm{p}}$ as defined in (\ref{eq:modeentropy}) cannot decrease, i.e.\ 
with $\dot{\mathcal{H}}_{\bm{p}} \equiv \mathrm{d}\mathcal{H}_{\bm{p}}/\mathrm{d}t$ it satisfies
\begin{equation}\label{eq:strongHtheorem}
    \dot{\mathcal{H}}_{\bm{p}}\geq0\,,
\end{equation}
for a large class of collision integrals $C[f](\bm{p})$ as specified in the following. The proof is a standard result going back to Boltzmann's original proof of the $H$-theorem in Ref.~\cite{Boltzmann:1872}\footnote{Boltzmann considered the ``classical" $H_\mathrm{cl}=-\int_{\bm p}f_{\bm p}\log f_{\bm p}$ instead of (\ref{eq:EB_entropy}), using a different sign convention.}.
Here we describe that not only the total $H$ including all entropy modes $\mathcal{H}_{\bm{p}}$ cannot decrease in standard kinetic theory, but in fact also each mode $\mathcal{H}_{\bm{p}}$ itself  cannot decrease, such that the strong form of the $H$-theorem as in (\ref{eq:strongHtheorem}) holds.
 
We start by considering the perturbative $2\leftrightarrow2$ collision integral for bosonic particles
\begin{equation}\label{eq:PertCollisionKernel}
    \begin{split}
    C^{2\leftrightarrow2}[f]({\bm{p}})&=\lambda^2\!\int_{\bm{q},\bm{r},\bm{s}}\bigg\{\underbrace{(f_{\bm{p}}+1)(f_{\bm{q}}+1)f_{\bm{r}}f_{\bm{s}}}_{\text{gain}}\\
    &\qquad\qquad\quad-\underbrace{f_{\bm{p}}f_{\bm{q}}(f_{\bm{r}}+1)(f_{\bm{s}}+1)}_{\text{loss}}\bigg\}\,,
\end{split}
\end{equation}
taking into account Bose enhancement, and the matrix element specified in terms of the coupling constant $\lambda$.
For the time derivative of the Boltzmann-Einstein entropy, defined in~(\ref{eq:EB_entropy}), one finds
\begin{equation}
\label{eq:dotH}
    \frac{\mathrm{d}H}{\mathrm{d}t}=\int_{\bm{p}}\frac{\mathrm{d}\mathcal{H}(f_{\bm{p}})}{\mathrm{d}t}=\int_{\bm{p}}\log(\frac{f_{\bm{p}}+1}{f_{\bm{p}}})\dot f_{\bm{p}}\equiv
    \int_{\bm{p}}\dot{\mathcal{H}}_{\bm{p}}\,,
\end{equation}
where $\dot f_{\bm{p}}\equiv\partial_t f(t,\bm{p})$. 
Using the Boltzmann equation (\ref{eq:BoltzmannEquation}) for the collision integral (\ref{eq:PertCollisionKernel}), we have
\[\begin{split}
    \frac{\mathrm{d}H}{\mathrm{d}t}&=\int_{\bm{p}}\log(\frac{f_{\bm{p}}+1}{f_{\bm{p}}})\dot f_{\bm{p}}\\
    &=\lambda^2\int_{\bm{p},\bm{q},\bm{r},\bm{s}}\log(\frac{f_{\bm{p}}+1}{f_{\bm{p}}})\bigg\{(f_{\bm{p}}+1)(f_{\bm{q}}+1)f_{\bm{r}}f_{\bm{s}}\\
    &\qquad\qquad\qquad-f_{\bm{p}}f_{\bm{q}}(f_{\bm{r}}+1)(f_{\bm{s}}+1)\bigg\}\,.
\end{split}\]
Copying the integrand four times, accordingly dividing by four, and renaming the integration variables in the second, third and fourth term according to ${\bm{p}}\leftrightarrow\bm{q},\bm{r}\leftrightarrow\bm{s}$; ${\bm{p}}\leftrightarrow\bm{r},\bm{q}\leftrightarrow\bm{s}$ and ${\bm{p}}\leftrightarrow\bm{s},\bm{q}\leftrightarrow\bm{r}$, respectively, we find
\begin{widetext}
\begin{equation}\begin{split}
    \frac{\mathrm{d}H}{\mathrm{d}t}&=\frac{\lambda^2}{4}\int_{\bm{p},\bm{q},\bm{r},\bm{s}}\bigg\{\log(\frac{f_{\bm{p}}+1}{f_{\bm{p}}})+\log(\frac{f_{\bm{q}}+1}{f_{\bm{q}}})-\log(\frac{f_{\bm{r}}+1}{f_{\bm{r}}})-\log(\frac{f_{\bm{s}}+1}{f_{\bm{s}}})\bigg\}\\
    &\qquad\qquad\quad\times\Big[(f_{\bm{p}}+1)(f_{\bm{q}}+1)f_{\bm{r}}f_{\bm{s}}-f_{\bm{p}}f_{\bm{q}}(f_{\bm{r}}+1)(f_{\bm{s}}+1)\Big]\\
    &=\frac{\lambda^2}{4}\int_{\bm{p},\bm{q},\bm{r},\bm{s}}\log(\frac{(f_{\bm{p}}+1)(f_{\bm{q}}+1)f_{\bm{r}}f_{\bm{s}}}{f_{\bm{p}}f_{\bm{q}}(f_{\bm{r}}+1)(f_{\bm{s}}+1)})\Big[(f_{\bm{p}}+1)(f_{\bm{q}}+1)f_{\bm{r}}f_{\bm{s}}-f_{\bm{p}}f_{\bm{q}}(f_{\bm{r}}+1)(f_{\bm{s}}+1)\Big]\\
    &\equiv\int_{\bm{p},\bm{q},\bm{r},\bm{s}}\dot{\mathscr{H}}(f_{\bm{p}},f_{\bm{q}},f_{\bm{r}},f_{\bm{s}})\,.
\end{split}
\label{eq:dotH2}
\end{equation}
\end{widetext}
In the last line we introduced
\[\begin{split}
    \dot{\mathscr{H}}&(f_{\bm{p}},f_{\bm{q}},f_{\bm{r}},f_{\bm{s}})=\frac{\lambda^2}{4}\log(\frac{(f_{\bm{p}}+1)(f_{\bm{q}}+1)f_{\bm{r}}f_{\bm{s}}}{f_{\bm{p}}f_{\bm{q}}(f_{\bm{r}}+1)(f_{\bm{s}}+1)})\\
    &\quad\times\Big[(f_{\bm{p}}+1)(f_{\bm{q}}+1)f_{\bm{r}}f_{\bm{s}}-f_{\bm{p}}f_{\bm{q}}(f_{\bm{r}}+1)(f_{\bm{s}}+1)\Big].
\end{split}\]
Noting that 
\[\log(\frac{a}{b})(a-b)\geq0\qquad\forall\,a,b>0\,,\]
with the equality only for $a=b$, one has
\begin{equation}\dot{\mathscr{H}}(f_{\bm{p}},f_{\bm{q}},f_{\bm{r}},f_{\bm{s}})\geq0\qquad\forall\,\bm{p,q,r,s}\,.
\end{equation}
Comparing the r.h.s.\ of (\ref{eq:dotH}) and (\ref{eq:dotH2}) we may write
\begin{equation}\label{eq:ModeEntropyChange}
    \dot{\mathcal{H}}_{\bm{p}} = \int_{\bm{q},\bm{r},\bm{s}}\dot{\mathscr{H}}(f_{\bm{p}},f_{\bm{q}},f_{\bm{r}},f_{\bm{s}})\,.
\end{equation}
Since for the mode entropy change 
in~(\ref{eq:ModeEntropyChange}) we are integrating a non-negative function $\dot{\mathscr{H}}(f_{\bm{p}},f_{\bm{q}},f_{\bm{r}},f_{\bm{s}})$, 
the mode entropy change is indeed non-negative, such that (\ref{eq:strongHtheorem}) holds.

This reasoning remains true even if we allow for momentum-dependent matrix elements $\mathcal{M}(\bm{p,q,r,s})$ rather than the coupling constants $\lambda$, such that the collision integral reads 
\begin{equation}\label{eq:NonPertCollisionKernel}
    \begin{split}
        C^{2\leftrightarrow2}&[f](\bm{p})=\int_{\bm{q,r,s}}|\mathcal{M}(\bm{p,q,r,s})|^2\\
    &\!\!\!\!\!\!\!\!\!\!\!\!\times\bigg\{\underbrace{(f_{\bm{p}}+1)(f_{\bm{q}}+1)f_{\bm{r}}f_{\bm{s}}}_{\text{gain}}-\underbrace{f_{\bm{p}}f_{\bm{q}}(f_{\bm{r}}+1)(f_{\bm{s}}+1)}_{\text{loss}}\bigg\}\,,
    \end{split}
\end{equation}
as long as the matrix elements $|\mathcal{M}(\bm{p},\bm{q},\bm{r},\bm{s})|^2$ obey the symmetries
\[\begin{split}
    |\mathcal{M}(\bm{p},\bm{q},\bm{r},\bm{s})|^2&=|\mathcal{M}(\bm{q},\bm{p},\bm{s},\bm{r})|^2\\
    &=|\mathcal{M}(\bm{r},\bm{s},\bm{p},\bm{q})|^2\\
    &=|\mathcal{M}(\bm{s},\bm{r},\bm{q},\bm{p})|^2\,.
\end{split}\]
This is satisfied for generalized Boltzmann equations such as the large-$N$ kinetic theory to next-to-leading order in $1/N$ considered in Ref.~\cite{Walz:2017ffj}.\\
\newpage
\bibliographystyle{apsrev4-1}
\bibliography{bibliography}

\begin{thebibliography}{37}%
\makeatletter
\providecommand \@ifxundefined [1]{%
 \@ifx{#1\undefined}
}%
\providecommand \@ifnum [1]{%
 \ifnum #1\expandafter \@firstoftwo
 \else \expandafter \@secondoftwo
 \fi
}%
\providecommand \@ifx [1]{%
 \ifx #1\expandafter \@firstoftwo
 \else \expandafter \@secondoftwo
 \fi
}%
\providecommand \natexlab [1]{#1}%
\providecommand \enquote  [1]{``#1''}%
\providecommand \bibnamefont  [1]{#1}%
\providecommand \bibfnamefont [1]{#1}%
\providecommand \citenamefont [1]{#1}%
\providecommand \href@noop [0]{\@secondoftwo}%
\providecommand \href [0]{\begingroup \@sanitize@url \@href}%
\providecommand \@href[1]{\@@startlink{#1}\@@href}%
\providecommand \@@href[1]{\endgroup#1\@@endlink}%
\providecommand \@sanitize@url [0]{\catcode `\\12\catcode `\$12\catcode `\&12\catcode `\#12\catcode `\^12\catcode `\_12\catcode `\%12\relax}%
\providecommand \@@startlink[1]{}%
\providecommand \@@endlink[0]{}%
\providecommand \url  [0]{\begingroup\@sanitize@url \@url }%
\providecommand \@url [1]{\endgroup\@href {#1}{\urlprefix }}%
\providecommand \urlprefix  [0]{URL }%
\providecommand \Eprint [0]{\href }%
\providecommand \doibase [0]{http://dx.doi.org/}%
\providecommand \selectlanguage [0]{\@gobble}%
\providecommand \bibinfo  [0]{\@secondoftwo}%
\providecommand \bibfield  [0]{\@secondoftwo}%
\providecommand \translation [1]{[#1]}%
\providecommand \BibitemOpen [0]{}%
\providecommand \bibitemStop [0]{}%
\providecommand \bibitemNoStop [0]{.\EOS\space}%
\providecommand \EOS [0]{\spacefactor3000\relax}%
\providecommand \BibitemShut  [1]{\csname bibitem#1\endcsname}%
\let\auto@bib@innerbib\@empty
%</preamble>
\bibitem [{\citenamefont {Feynman}\ \emph {et~al.}(2011)\citenamefont {Feynman}, \citenamefont {Leighton},\ and\ \citenamefont {Sands}}]{Feynman:2011}%
  \BibitemOpen
  \bibfield  {author} {\bibinfo {author} {\bibfnamefont {R.~P.}\ \bibnamefont {Feynman}}, \bibinfo {author} {\bibfnamefont {R.~B.}\ \bibnamefont {Leighton}}, \ and\ \bibinfo {author} {\bibfnamefont {M.}~\bibnamefont {Sands}},\ }\href@noop {} {\emph {\bibinfo {title} {The Feynman Lectures on Physics, Volume I: Mainly Mechanics, Radiation, and Heat}}},\ \bibinfo {edition} {new millennium edition}\ ed.\ (\bibinfo  {publisher} {Basic Books},\ \bibinfo {address} {New York},\ \bibinfo {year} {2011})\ \bibinfo {note} {\url{www.feynmanlectures.caltech.edu/I_46.html}}\BibitemShut {NoStop}%
\bibitem [{\citenamefont {Penrose}(1979)}]{Penrose:1979}%
  \BibitemOpen
  \bibfield  {author} {\bibinfo {author} {\bibfnamefont {R.}~\bibnamefont {Penrose}},\ }in\ \href@noop {} {\emph {\bibinfo {booktitle} {General Relativity: An Einstein Centenary Survey}}}\ (\bibinfo  {publisher} {Cambridge University Press},\ \bibinfo {address} {Cambridge},\ \bibinfo {year} {1979})\ pp.\ \bibinfo {pages} {581--638}\BibitemShut {NoStop}%
\bibitem [{\citenamefont {Berges}(2002)}]{Berges:2001fi}%
  \BibitemOpen
  \bibfield  {author} {\bibinfo {author} {\bibfnamefont {J.}~\bibnamefont {Berges}},\ }\href {\doibase 10.1016/S0375-9474(01)01295-7} {\bibfield  {journal} {\bibinfo  {journal} {Nucl. Phys. A}\ }\textbf {\bibinfo {volume} {699}},\ \bibinfo {pages} {847} (\bibinfo {year} {2002})},\ \Eprint {http://arxiv.org/abs/hep-ph/0105311} {arXiv:hep-ph/0105311} \BibitemShut {NoStop}%
\bibitem [{\citenamefont {Berges}\ \emph {et~al.}(2008)\citenamefont {Berges}, \citenamefont {Rothkopf},\ and\ \citenamefont {Schmidt}}]{Berges:2008wm}%
  \BibitemOpen
  \bibfield  {author} {\bibinfo {author} {\bibfnamefont {J.}~\bibnamefont {Berges}}, \bibinfo {author} {\bibfnamefont {A.}~\bibnamefont {Rothkopf}}, \ and\ \bibinfo {author} {\bibfnamefont {J.}~\bibnamefont {Schmidt}},\ }\href {\doibase 10.1103/PhysRevLett.101.041603} {\bibfield  {journal} {\bibinfo  {journal} {Phys. Rev. Lett.}\ }\textbf {\bibinfo {volume} {101}},\ \bibinfo {pages} {041603} (\bibinfo {year} {2008})},\ \Eprint {http://arxiv.org/abs/0803.0131} {arXiv:0803.0131 [hep-ph]} \BibitemShut {NoStop}%
\bibitem [{\citenamefont {Pi{\~n}eiro~Orioli}\ \emph {et~al.}(2015)\citenamefont {Pi{\~n}eiro~Orioli}, \citenamefont {Boguslavski},\ and\ \citenamefont {Berges}}]{Orioli:2015cpb}%
  \BibitemOpen
  \bibfield  {author} {\bibinfo {author} {\bibfnamefont {A.}~\bibnamefont {Pi{\~n}eiro~Orioli}}, \bibinfo {author} {\bibfnamefont {K.}~\bibnamefont {Boguslavski}}, \ and\ \bibinfo {author} {\bibfnamefont {J.}~\bibnamefont {Berges}},\ }\href {\doibase 10.1103/PhysRevD.92.025041} {\bibfield  {journal} {\bibinfo  {journal} {Phys. Rev. D}\ }\textbf {\bibinfo {volume} {92}},\ \bibinfo {pages} {025041} (\bibinfo {year} {2015})},\ \Eprint {http://arxiv.org/abs/1503.02498} {arXiv:1503.02498 [hep-ph]} \BibitemShut {NoStop}%
\bibitem [{\citenamefont {Chantesana}\ \emph {et~al.}(2019)\citenamefont {Chantesana}, \citenamefont {Pi{\~n}eiro~Orioli},\ and\ \citenamefont {Gasenzer}}]{Chantesana:2018qsb}%
  \BibitemOpen
  \bibfield  {author} {\bibinfo {author} {\bibfnamefont {I.}~\bibnamefont {Chantesana}}, \bibinfo {author} {\bibfnamefont {A.}~\bibnamefont {Pi{\~n}eiro~Orioli}}, \ and\ \bibinfo {author} {\bibfnamefont {T.}~\bibnamefont {Gasenzer}},\ }\href {\doibase 10.1103/PhysRevA.99.043620} {\bibfield  {journal} {\bibinfo  {journal} {Phys. Rev. A}\ }\textbf {\bibinfo {volume} {99}},\ \bibinfo {pages} {043620} (\bibinfo {year} {2019})},\ \Eprint {http://arxiv.org/abs/1801.09490} {arXiv:1801.09490 [cond-mat.quant-gas]} \BibitemShut {NoStop}%
\bibitem [{\citenamefont {Berges}\ \emph {et~al.}(2021)\citenamefont {Berges}, \citenamefont {Heller}, \citenamefont {Mazeliauskas},\ and\ \citenamefont {Venugopalan}}]{Berges:2020fwq}%
  \BibitemOpen
  \bibfield  {author} {\bibinfo {author} {\bibfnamefont {J.}~\bibnamefont {Berges}}, \bibinfo {author} {\bibfnamefont {M.~P.}\ \bibnamefont {Heller}}, \bibinfo {author} {\bibfnamefont {A.}~\bibnamefont {Mazeliauskas}}, \ and\ \bibinfo {author} {\bibfnamefont {R.}~\bibnamefont {Venugopalan}},\ }\href {\doibase 10.1103/RevModPhys.93.035003} {\bibfield  {journal} {\bibinfo  {journal} {Rev. Mod. Phys.}\ }\textbf {\bibinfo {volume} {93}},\ \bibinfo {pages} {035003} (\bibinfo {year} {2021})},\ \Eprint {http://arxiv.org/abs/2005.12299} {arXiv:2005.12299 [hep-th]} \BibitemShut {NoStop}%
\bibitem [{\citenamefont {Pr{\"u}fer}\ \emph {et~al.}(2018)\citenamefont {Pr{\"u}fer}, \citenamefont {Kunkel}, \citenamefont {Strobel}, \citenamefont {Lannig}, \citenamefont {Linnemann}, \citenamefont {Schmied}, \citenamefont {Berges}, \citenamefont {Gasenzer},\ and\ \citenamefont {Oberthaler}}]{Prufer:2018hto}%
  \BibitemOpen
  \bibfield  {author} {\bibinfo {author} {\bibfnamefont {M.}~\bibnamefont {Pr{\"u}fer}}, \bibinfo {author} {\bibfnamefont {P.}~\bibnamefont {Kunkel}}, \bibinfo {author} {\bibfnamefont {H.}~\bibnamefont {Strobel}}, \bibinfo {author} {\bibfnamefont {S.}~\bibnamefont {Lannig}}, \bibinfo {author} {\bibfnamefont {D.}~\bibnamefont {Linnemann}}, \bibinfo {author} {\bibfnamefont {C.-M.}\ \bibnamefont {Schmied}}, \bibinfo {author} {\bibfnamefont {J.}~\bibnamefont {Berges}}, \bibinfo {author} {\bibfnamefont {T.}~\bibnamefont {Gasenzer}}, \ and\ \bibinfo {author} {\bibfnamefont {M.~K.}\ \bibnamefont {Oberthaler}},\ }\href {\doibase 10.1038/s41586-018-0659-0} {\bibfield  {journal} {\bibinfo  {journal} {Nature}\ }\textbf {\bibinfo {volume} {563}},\ \bibinfo {pages} {217} (\bibinfo {year} {2018})},\ \Eprint {http://arxiv.org/abs/1805.11881} {arXiv:1805.11881 [cond-mat.quant-gas]} \BibitemShut {NoStop}%
\bibitem [{\citenamefont {Erne}\ \emph {et~al.}(2018)\citenamefont {Erne}, \citenamefont {B{\"u}cker}, \citenamefont {Gasenzer}, \citenamefont {Berges},\ and\ \citenamefont {Schmiedmayer}}]{Erne:2018gmz}%
  \BibitemOpen
  \bibfield  {author} {\bibinfo {author} {\bibfnamefont {S.}~\bibnamefont {Erne}}, \bibinfo {author} {\bibfnamefont {R.}~\bibnamefont {B{\"u}cker}}, \bibinfo {author} {\bibfnamefont {T.}~\bibnamefont {Gasenzer}}, \bibinfo {author} {\bibfnamefont {J.}~\bibnamefont {Berges}}, \ and\ \bibinfo {author} {\bibfnamefont {J.}~\bibnamefont {Schmiedmayer}},\ }\href {\doibase 10.1038/s41586-018-0667-0} {\bibfield  {journal} {\bibinfo  {journal} {Nature}\ }\textbf {\bibinfo {volume} {563}},\ \bibinfo {pages} {225} (\bibinfo {year} {2018})},\ \Eprint {http://arxiv.org/abs/1805.12310} {arXiv:1805.12310 [cond-mat.quant-gas]} \BibitemShut {NoStop}%
\bibitem [{\citenamefont {Glidden}\ \emph {et~al.}(2021)\citenamefont {Glidden}, \citenamefont {Eigen}, \citenamefont {Dogra}, \citenamefont {Hilker}, \citenamefont {Smith},\ and\ \citenamefont {Hadzibabic}}]{Glidden:2020qmu}%
  \BibitemOpen
  \bibfield  {author} {\bibinfo {author} {\bibfnamefont {J.~A.~P.}\ \bibnamefont {Glidden}}, \bibinfo {author} {\bibfnamefont {C.}~\bibnamefont {Eigen}}, \bibinfo {author} {\bibfnamefont {L.~H.}\ \bibnamefont {Dogra}}, \bibinfo {author} {\bibfnamefont {T.~A.}\ \bibnamefont {Hilker}}, \bibinfo {author} {\bibfnamefont {R.~P.}\ \bibnamefont {Smith}}, \ and\ \bibinfo {author} {\bibfnamefont {Z.}~\bibnamefont {Hadzibabic}},\ }\href {\doibase 10.1038/s41567-020-01114-x} {\bibfield  {journal} {\bibinfo  {journal} {Nature Phys.}\ }\textbf {\bibinfo {volume} {17}},\ \bibinfo {pages} {457} (\bibinfo {year} {2021})},\ \Eprint {http://arxiv.org/abs/2006.01118} {arXiv:2006.01118 [cond-mat.quant-gas]} \BibitemShut {NoStop}%
\bibitem [{\citenamefont {Gazo}\ \emph {et~al.}(2025)\citenamefont {Gazo}, \citenamefont {Karailiev}, \citenamefont {Satoor}, \citenamefont {Eigen}, \citenamefont {Ga{\l}ka},\ and\ \citenamefont {Hadzibabic}}]{Gazo:2023exc}%
  \BibitemOpen
  \bibfield  {author} {\bibinfo {author} {\bibfnamefont {M.}~\bibnamefont {Gazo}}, \bibinfo {author} {\bibfnamefont {A.}~\bibnamefont {Karailiev}}, \bibinfo {author} {\bibfnamefont {T.}~\bibnamefont {Satoor}}, \bibinfo {author} {\bibfnamefont {C.}~\bibnamefont {Eigen}}, \bibinfo {author} {\bibfnamefont {M.}~\bibnamefont {Ga{\l}ka}}, \ and\ \bibinfo {author} {\bibfnamefont {Z.}~\bibnamefont {Hadzibabic}},\ }\href {\doibase 10.1126/science.ado3487} {\bibfield  {journal} {\bibinfo  {journal} {Science}\ }\textbf {\bibinfo {volume} {389}},\ \bibinfo {pages} {ado3487} (\bibinfo {year} {2025})},\ \Eprint {http://arxiv.org/abs/2312.09248} {arXiv:2312.09248 [cond-mat.quant-gas]} \BibitemShut {NoStop}%
\bibitem [{\citenamefont {Huh}\ \emph {et~al.}(2024)\citenamefont {Huh}, \citenamefont {Mukherjee}, \citenamefont {Kwon}, \citenamefont {Seo}, \citenamefont {Hur}, \citenamefont {Mistakidis}, \citenamefont {Sadeghpour},\ and\ \citenamefont {Choi}}]{Huh:2023xso}%
  \BibitemOpen
  \bibfield  {author} {\bibinfo {author} {\bibfnamefont {S.}~\bibnamefont {Huh}}, \bibinfo {author} {\bibfnamefont {K.}~\bibnamefont {Mukherjee}}, \bibinfo {author} {\bibfnamefont {K.}~\bibnamefont {Kwon}}, \bibinfo {author} {\bibfnamefont {J.}~\bibnamefont {Seo}}, \bibinfo {author} {\bibfnamefont {J.}~\bibnamefont {Hur}}, \bibinfo {author} {\bibfnamefont {S.~I.}\ \bibnamefont {Mistakidis}}, \bibinfo {author} {\bibfnamefont {H.~R.}\ \bibnamefont {Sadeghpour}}, \ and\ \bibinfo {author} {\bibfnamefont {J.-y.}\ \bibnamefont {Choi}},\ }\href {\doibase 10.1038/s41567-023-02339-2} {\bibfield  {journal} {\bibinfo  {journal} {Nature Phys.}\ }\textbf {\bibinfo {volume} {20}},\ \bibinfo {pages} {402} (\bibinfo {year} {2024})},\ \Eprint {http://arxiv.org/abs/2303.05230} {arXiv:2303.05230 [cond-mat.quant-gas]} \BibitemShut {NoStop}%
\bibitem [{\citenamefont {Svistunov}(1991)}]{Svistunov:1991}%
  \BibitemOpen
  \bibfield  {author} {\bibinfo {author} {\bibfnamefont {B.~V.}\ \bibnamefont {Svistunov}},\ }\href@noop {} {\bibfield  {journal} {\bibinfo  {journal} {J. Moscow Phys. Soc}\ }\textbf {\bibinfo {volume} {1}},\ \bibinfo {pages} {373} (\bibinfo {year} {1991})}\BibitemShut {NoStop}%
\bibitem [{\citenamefont {Bray}(1994)}]{Bray:1994zz}%
  \BibitemOpen
  \bibfield  {author} {\bibinfo {author} {\bibfnamefont {A.~J.}\ \bibnamefont {Bray}},\ }\href {\doibase 10.1080/00018739400101505} {\bibfield  {journal} {\bibinfo  {journal} {Adv. Phys.}\ }\textbf {\bibinfo {volume} {43}},\ \bibinfo {pages} {357} (\bibinfo {year} {1994})},\ \Eprint {http://arxiv.org/abs/cond-mat/9501089} {arXiv:cond-mat/9501089} \BibitemShut {NoStop}%
\bibitem [{\citenamefont {Berloff}\ and\ \citenamefont {Svistunov}(2002)}]{Berloff:2002}%
  \BibitemOpen
  \bibfield  {author} {\bibinfo {author} {\bibfnamefont {N.~G.}\ \bibnamefont {Berloff}}\ and\ \bibinfo {author} {\bibfnamefont {B.~V.}\ \bibnamefont {Svistunov}},\ }\href@noop {} {\bibfield  {journal} {\bibinfo  {journal} {Phys. Rev. A}\ }\textbf {\bibinfo {volume} {66}},\ \bibinfo {pages} {013603} (\bibinfo {year} {2002})}\BibitemShut {NoStop}%
\bibitem [{\citenamefont {Berges}\ and\ \citenamefont {Sexty}(2012)}]{Berges:2012us}%
  \BibitemOpen
  \bibfield  {author} {\bibinfo {author} {\bibfnamefont {J.}~\bibnamefont {Berges}}\ and\ \bibinfo {author} {\bibfnamefont {D.}~\bibnamefont {Sexty}},\ }\href {\doibase 10.1103/PhysRevLett.108.161601} {\bibfield  {journal} {\bibinfo  {journal} {Phys. Rev. Lett.}\ }\textbf {\bibinfo {volume} {108}},\ \bibinfo {pages} {161601} (\bibinfo {year} {2012})},\ \Eprint {http://arxiv.org/abs/1201.0687} {arXiv:1201.0687 [hep-ph]} \BibitemShut {NoStop}%
\bibitem [{\citenamefont {Nowak}\ and\ \citenamefont {Gasenzer}(2014)}]{Nowak:2012gd}%
  \BibitemOpen
  \bibfield  {author} {\bibinfo {author} {\bibfnamefont {B.}~\bibnamefont {Nowak}}\ and\ \bibinfo {author} {\bibfnamefont {T.}~\bibnamefont {Gasenzer}},\ }\href {\doibase 10.1088/1367-2630/16/9/093052} {\bibfield  {journal} {\bibinfo  {journal} {New J. Phys.}\ }\textbf {\bibinfo {volume} {16}},\ \bibinfo {pages} {093052} (\bibinfo {year} {2014})},\ \Eprint {http://arxiv.org/abs/1206.3181} {arXiv:1206.3181 [cond-mat.quant-gas]} \BibitemShut {NoStop}%
\bibitem [{\citenamefont {Preis}\ \emph {et~al.}(2023)\citenamefont {Preis}, \citenamefont {Heller},\ and\ \citenamefont {Berges}}]{Preis:2022uqs}%
  \BibitemOpen
  \bibfield  {author} {\bibinfo {author} {\bibfnamefont {T.}~\bibnamefont {Preis}}, \bibinfo {author} {\bibfnamefont {M.~P.}\ \bibnamefont {Heller}}, \ and\ \bibinfo {author} {\bibfnamefont {J.}~\bibnamefont {Berges}},\ }\href {\doibase 10.1103/PhysRevLett.130.031602} {\bibfield  {journal} {\bibinfo  {journal} {Phys. Rev. Lett.}\ }\textbf {\bibinfo {volume} {130}},\ \bibinfo {pages} {031602} (\bibinfo {year} {2023})},\ \Eprint {http://arxiv.org/abs/2209.14883} {arXiv:2209.14883 [hep-ph]} \BibitemShut {NoStop}%
\bibitem [{\citenamefont {Nazarenko}(2011)}]{Nazarenko:2011}%
  \BibitemOpen
  \bibfield  {author} {\bibinfo {author} {\bibfnamefont {S.}~\bibnamefont {Nazarenko}},\ }\href {\doibase 10.1007/978-3-642-15942-8} {\bibfield  {journal} {\bibinfo  {journal} {Springer-Verlag Berlin Heidelberg}\ } (\bibinfo {year} {2011}),\ 10.1007/978-3-642-15942-8}\BibitemShut {NoStop}%
\bibitem [{\citenamefont {Micha}\ and\ \citenamefont {Tkachev}(2004)}]{Micha:2004bv}%
  \BibitemOpen
  \bibfield  {author} {\bibinfo {author} {\bibfnamefont {R.}~\bibnamefont {Micha}}\ and\ \bibinfo {author} {\bibfnamefont {I.~I.}\ \bibnamefont {Tkachev}},\ }\href {\doibase 10.1103/PhysRevD.70.043538} {\bibfield  {journal} {\bibinfo  {journal} {Phys. Rev. D}\ }\textbf {\bibinfo {volume} {70}},\ \bibinfo {pages} {043538} (\bibinfo {year} {2004})},\ \Eprint {http://arxiv.org/abs/hep-ph/0403101} {arXiv:hep-ph/0403101} \BibitemShut {NoStop}%
\bibitem [{\citenamefont {Berges}\ \emph {et~al.}(2015)\citenamefont {Berges}, \citenamefont {Boguslavski}, \citenamefont {Schlichting},\ and\ \citenamefont {Venugopalan}}]{Berges:2014bba}%
  \BibitemOpen
  \bibfield  {author} {\bibinfo {author} {\bibfnamefont {J.}~\bibnamefont {Berges}}, \bibinfo {author} {\bibfnamefont {K.}~\bibnamefont {Boguslavski}}, \bibinfo {author} {\bibfnamefont {S.}~\bibnamefont {Schlichting}}, \ and\ \bibinfo {author} {\bibfnamefont {R.}~\bibnamefont {Venugopalan}},\ }\href {\doibase 10.1103/PhysRevLett.114.061601} {\bibfield  {journal} {\bibinfo  {journal} {Phys. Rev. Lett.}\ }\textbf {\bibinfo {volume} {114}},\ \bibinfo {pages} {061601} (\bibinfo {year} {2015})},\ \Eprint {http://arxiv.org/abs/1408.1670} {arXiv:1408.1670 [hep-ph]} \BibitemShut {NoStop}%
\bibitem [{\citenamefont {Stamper-Kurn}\ and\ \citenamefont {Ueda}(2013)}]{Stamper-Kurn:2012bxy}%
  \BibitemOpen
  \bibfield  {author} {\bibinfo {author} {\bibfnamefont {D.~M.}\ \bibnamefont {Stamper-Kurn}}\ and\ \bibinfo {author} {\bibfnamefont {M.}~\bibnamefont {Ueda}},\ }\href {\doibase 10.1103/RevModPhys.85.1191} {\bibfield  {journal} {\bibinfo  {journal} {Rev. Mod. Phys.}\ }\textbf {\bibinfo {volume} {85}},\ \bibinfo {pages} {1191} (\bibinfo {year} {2013})},\ \Eprint {http://arxiv.org/abs/1205.1888} {arXiv:1205.1888 [cond-mat.quant-gas]} \BibitemShut {NoStop}%
\bibitem [{\citenamefont {Pr\"ufer}\ \emph {et~al.}(2020)\citenamefont {Pr\"ufer}, \citenamefont {Zache}, \citenamefont {Kunkel}, \citenamefont {Lannig}, \citenamefont {Bonnin}, \citenamefont {Strobel}, \citenamefont {Berges},\ and\ \citenamefont {Oberthaler}}]{Prufer:2019kak}%
  \BibitemOpen
  \bibfield  {author} {\bibinfo {author} {\bibfnamefont {M.}~\bibnamefont {Pr\"ufer}}, \bibinfo {author} {\bibfnamefont {T.~V.}\ \bibnamefont {Zache}}, \bibinfo {author} {\bibfnamefont {P.}~\bibnamefont {Kunkel}}, \bibinfo {author} {\bibfnamefont {S.}~\bibnamefont {Lannig}}, \bibinfo {author} {\bibfnamefont {A.}~\bibnamefont {Bonnin}}, \bibinfo {author} {\bibfnamefont {H.}~\bibnamefont {Strobel}}, \bibinfo {author} {\bibfnamefont {J.}~\bibnamefont {Berges}}, \ and\ \bibinfo {author} {\bibfnamefont {M.~K.}\ \bibnamefont {Oberthaler}},\ }\href {\doibase 10.1038/s41567-020-0933-6} {\bibfield  {journal} {\bibinfo  {journal} {Nature Phys.}\ }\textbf {\bibinfo {volume} {16}},\ \bibinfo {pages} {1012} (\bibinfo {year} {2020})},\ \Eprint {http://arxiv.org/abs/1909.05120} {arXiv:1909.05120 [cond-mat.quant-gas]} \BibitemShut {NoStop}%
\bibitem [{\citenamefont {Einstein}(1925)}]{Einstein:1925}%
  \BibitemOpen
  \bibfield  {author} {\bibinfo {author} {\bibfnamefont {A.}~\bibnamefont {Einstein}},\ }\href@noop {} {\bibfield  {journal} {\bibinfo  {journal} {Sitzungsberichte der Preussischen Akademie der Wissenschaften}\ ,\ \bibinfo {pages} {3}} (\bibinfo {year} {1925})},\ \bibinfo {note} {eq. (29a), p.~3}\BibitemShut {NoStop}%
\bibitem [{\citenamefont {Shannon}(1948)}]{Shannon:1948}%
  \BibitemOpen
  \bibfield  {author} {\bibinfo {author} {\bibfnamefont {C.~E.}\ \bibnamefont {Shannon}},\ }\href@noop {} {\bibfield  {journal} {\bibinfo  {journal} {Bell System Technical Journal}\ }\textbf {\bibinfo {volume} {27}},\ \bibinfo {pages} {379} (\bibinfo {year} {1948})}\BibitemShut {NoStop}%
\bibitem [{\citenamefont {Boltzmann}(1872)}]{Boltzmann:1872}%
  \BibitemOpen
  \bibfield  {author} {\bibinfo {author} {\bibfnamefont {L.}~\bibnamefont {Boltzmann}},\ }\href@noop {} {\bibfield  {journal} {\bibinfo  {journal} {Sitzungsberichte der Akademie der Wissenschaften Wien, II}\ }\textbf {\bibinfo {volume} {66}},\ \bibinfo {pages} {275} (\bibinfo {year} {1872})}\BibitemShut {NoStop}%
\bibitem [{\citenamefont {Tolman}(1938)}]{Tolman:1938}%
  \BibitemOpen
  \bibfield  {author} {\bibinfo {author} {\bibfnamefont {R.~C.}\ \bibnamefont {Tolman}},\ }\href@noop {} {\emph {\bibinfo {title} {The Principles of Statistical Mechanics}}}\ (\bibinfo  {publisher} {Oxford University Press, Clarendon Press},\ \bibinfo {address} {Oxford},\ \bibinfo {year} {1938})\ \bibinfo {note} {chapter XII, Eq. (103.13)}\BibitemShut {NoStop}%
\bibitem [{\citenamefont {Berges}\ \emph {et~al.}(2018)\citenamefont {Berges}, \citenamefont {Floerchinger},\ and\ \citenamefont {Venugopalan}}]{Berges:2017hne}%
  \BibitemOpen
  \bibfield  {author} {\bibinfo {author} {\bibfnamefont {J.}~\bibnamefont {Berges}}, \bibinfo {author} {\bibfnamefont {S.}~\bibnamefont {Floerchinger}}, \ and\ \bibinfo {author} {\bibfnamefont {R.}~\bibnamefont {Venugopalan}},\ }\href {\doibase 10.1007/JHEP04(2018)145} {\bibfield  {journal} {\bibinfo  {journal} {JHEP}\ }\textbf {\bibinfo {volume} {04}},\ \bibinfo {pages} {145} (\bibinfo {year} {2018})},\ \Eprint {http://arxiv.org/abs/1712.09362} {arXiv:1712.09362 [hep-th]} \BibitemShut {NoStop}%
\bibitem [{\citenamefont {Kunkel}\ \emph {et~al.}(2019)\citenamefont {Kunkel}, \citenamefont {Pr{\"u}fer}, \citenamefont {Lannig}, \citenamefont {Rosa-Medina}, \citenamefont {Bonnin}, \citenamefont {G{\"a}rttner}, \citenamefont {Strobel},\ and\ \citenamefont {Oberthaler}}]{Kunkel:2019vud}%
  \BibitemOpen
  \bibfield  {author} {\bibinfo {author} {\bibfnamefont {P.}~\bibnamefont {Kunkel}}, \bibinfo {author} {\bibfnamefont {M.}~\bibnamefont {Pr{\"u}fer}}, \bibinfo {author} {\bibfnamefont {S.}~\bibnamefont {Lannig}}, \bibinfo {author} {\bibfnamefont {R.}~\bibnamefont {Rosa-Medina}}, \bibinfo {author} {\bibfnamefont {A.}~\bibnamefont {Bonnin}}, \bibinfo {author} {\bibfnamefont {M.}~\bibnamefont {G{\"a}rttner}}, \bibinfo {author} {\bibfnamefont {H.}~\bibnamefont {Strobel}}, \ and\ \bibinfo {author} {\bibfnamefont {M.~K.}\ \bibnamefont {Oberthaler}},\ }\href {\doibase 10.1103/PhysRevLett.123.063603} {\bibfield  {journal} {\bibinfo  {journal} {Phys. Rev. Lett.}\ }\textbf {\bibinfo {volume} {123}},\ \bibinfo {pages} {063603} (\bibinfo {year} {2019})},\ \Eprint {http://arxiv.org/abs/1904.01471} {arXiv:1904.01471 [cond-mat.quant-gas]} \BibitemShut {NoStop}%
\bibitem [{\citenamefont {Berges}(2015)}]{berges2015}%
  \BibitemOpen
  \bibfield  {author} {\bibinfo {author} {\bibfnamefont {J.}~\bibnamefont {Berges}},\ }\href {https://arxiv.org/abs/1503.02907} {\enquote {\bibinfo {title} {Nonequilibrium quantum fields: From cold atoms to cosmology},}\ } (\bibinfo {year} {2015}),\ \Eprint {http://arxiv.org/abs/1503.02907} {arXiv:1503.02907 [hep-ph]} \BibitemShut {NoStop}%
\bibitem [{\citenamefont {Ahmed}\ and\ \citenamefont {Gokhale}(1989)}]{30996}%
  \BibitemOpen
  \bibfield  {author} {\bibinfo {author} {\bibfnamefont {N.}~\bibnamefont {Ahmed}}\ and\ \bibinfo {author} {\bibfnamefont {D.}~\bibnamefont {Gokhale}},\ }\href {\doibase 10.1109/18.30996} {\bibfield  {journal} {\bibinfo  {journal} {IEEE Transactions on Information Theory}\ }\textbf {\bibinfo {volume} {35}},\ \bibinfo {pages} {688} (\bibinfo {year} {1989})}\BibitemShut {NoStop}%
\bibitem [{\citenamefont {Steeg}(2014)}]{versteeg2014}%
  \BibitemOpen
  \bibfield  {author} {\bibinfo {author} {\bibfnamefont {G.~V.}\ \bibnamefont {Steeg}},\ }\href@noop {} {\enquote {\bibinfo {title} {Non-parametric entropy estimation toolbox (npeet)},}\ }\bibinfo {howpublished} {\url{github.com/gregversteeg/NPEET}} (\bibinfo {year} {2014}),\ \bibinfo {note} {gitHub repository}\BibitemShut {NoStop}%
\bibitem [{\citenamefont {Deller}\ \emph {et~al.}(2025)\citenamefont {Deller}, \citenamefont {Gärttner}, \citenamefont {Haas}, \citenamefont {Oberthaler}, \citenamefont {Reh},\ and\ \citenamefont {Strobel}}]{Deller_2025}%
  \BibitemOpen
  \bibfield  {author} {\bibinfo {author} {\bibfnamefont {Y.}~\bibnamefont {Deller}}, \bibinfo {author} {\bibfnamefont {M.}~\bibnamefont {Gärttner}}, \bibinfo {author} {\bibfnamefont {T.}~\bibnamefont {Haas}}, \bibinfo {author} {\bibfnamefont {M.~K.}\ \bibnamefont {Oberthaler}}, \bibinfo {author} {\bibfnamefont {M.}~\bibnamefont {Reh}}, \ and\ \bibinfo {author} {\bibfnamefont {H.}~\bibnamefont {Strobel}},\ }\href {\doibase 10.1088/1367-2630/adc32a} {\bibfield  {journal} {\bibinfo  {journal} {New Journal of Physics}\ }\textbf {\bibinfo {volume} {27}},\ \bibinfo {pages} {043004} (\bibinfo {year} {2025})}\BibitemShut {NoStop}%
\bibitem [{\citenamefont {Kozachenko}\ and\ \citenamefont {Leonenko}(1987)}]{kozachenko1987}%
  \BibitemOpen
  \bibfield  {author} {\bibinfo {author} {\bibfnamefont {L.~F.}\ \bibnamefont {Kozachenko}}\ and\ \bibinfo {author} {\bibfnamefont {N.~N.}\ \bibnamefont {Leonenko}},\ }\href@noop {} {\bibfield  {journal} {\bibinfo  {journal} {Problemy Peredachi Informatsii}\ }\textbf {\bibinfo {volume} {23}},\ \bibinfo {pages} {9} (\bibinfo {year} {1987})}\BibitemShut {NoStop}%
\bibitem [{\citenamefont {Shen}\ and\ \citenamefont {Berges}(2020)}]{Shen_2020}%
  \BibitemOpen
  \bibfield  {author} {\bibinfo {author} {\bibfnamefont {L.}~\bibnamefont {Shen}}\ and\ \bibinfo {author} {\bibfnamefont {J.}~\bibnamefont {Berges}},\ }\href {\doibase 10.1103/physrevd.101.056009} {\bibfield  {journal} {\bibinfo  {journal} {Physical Review D}\ }\textbf {\bibinfo {volume} {101}} (\bibinfo {year} {2020}),\ 10.1103/physrevd.101.056009}\BibitemShut {NoStop}%
\bibitem [{\citenamefont {Uehling}\ and\ \citenamefont {Uhlenbeck}(1933)}]{uehling1933transport}%
  \BibitemOpen
  \bibfield  {author} {\bibinfo {author} {\bibfnamefont {E.~A.}\ \bibnamefont {Uehling}}\ and\ \bibinfo {author} {\bibfnamefont {G.~E.}\ \bibnamefont {Uhlenbeck}},\ }\href {\doibase 10.1103/PhysRev.43.552} {\bibfield  {journal} {\bibinfo  {journal} {Physical Review}\ }\textbf {\bibinfo {volume} {43}},\ \bibinfo {pages} {552} (\bibinfo {year} {1933})}\BibitemShut {NoStop}%
\bibitem [{\citenamefont {Walz}\ \emph {et~al.}(2018)\citenamefont {Walz}, \citenamefont {Boguslavski},\ and\ \citenamefont {Berges}}]{Walz:2017ffj}%
  \BibitemOpen
  \bibfield  {author} {\bibinfo {author} {\bibfnamefont {R.}~\bibnamefont {Walz}}, \bibinfo {author} {\bibfnamefont {K.}~\bibnamefont {Boguslavski}}, \ and\ \bibinfo {author} {\bibfnamefont {J.}~\bibnamefont {Berges}},\ }\href {\doibase 10.1103/PhysRevD.97.116011} {\bibfield  {journal} {\bibinfo  {journal} {Phys. Rev. D}\ }\textbf {\bibinfo {volume} {97}},\ \bibinfo {pages} {116011} (\bibinfo {year} {2018})},\ \Eprint {http://arxiv.org/abs/1710.11146} {arXiv:1710.11146 [hep-ph]} \BibitemShut {NoStop}%
\end{thebibliography}%

\end{document}